\RequirePackage[hyphens]{url}
\documentclass[titlecompat,tight,twocolumn,10pt]{article}
\usepackage{usenix}

\PassOptionsToPackage{hyphens}{url}
\PassOptionsToPackage{breaklinks}{hyperref}
\usepackage{adjustbox}
\usepackage{xspace}
\usepackage{fullwidth}
\usepackage{graphicx}
\usepackage{enumitem}
\usepackage{booktabs}
\usepackage{multirow}
\usepackage[binary-units=true]{siunitx}
\usepackage{amsfonts,amsthm,amsmath}
\usepackage{subcaption}

\usepackage{color}
\usepackage[ruled,vlined,linesnumbered,noend]{algorithm2e}
\SetKw{Continue}{continue}
\SetKw{Break}{break}
\SetKw{None}{None}
\SetKw{len}{len}
\SetKw{append}{append}
\SetKw{true}{True}
\SetKw{union}{union}
\SetKw{random}{random}
\SetKw{sample}{sample}
\SetKw{forrr}{for}
\SetKw{ifff}{if}
\usepackage{cleveref}
\usepackage{tabularx} 
\usepackage{makecell}
\usepackage{algorithmic}

\usepackage{authblk}
\DeclareMathAlphabet{\mathsc}{OT1}{cmr}{m}{sc}
\newcommand\sarah[1]{\textcolor{red}{Sarah: #1}}
\newcommand\h[1]{\textcolor{magenta}{Haaroon: #1}}
\newcommand\george[1]{\textcolor{green}{George: #1}}
\newcommand\rainer[1]{\textcolor{blue}{Rainer: #1}}
\newcommand\bernhard[1]{\textcolor{yellow}{Bernhard: #1}}
\newcommand\sofia[1]{\textcolor{blue}{Sofia: #1}}
\renewcommand\sarah[1]{}
\renewcommand\h[1]{}
\renewcommand\george[1]{}
\renewcommand\rainer[1]{}
\renewcommand\bernhard[1]{}
\renewcommand\sofia[1]{}

\newcommand{\Cpp}{C\nolinebreak\hspace{-.05em}%
  \raisebox{.4ex}{\tiny\bf +}\nolinebreak\hspace{-.10em}%
  \raisebox{.4ex}{\tiny\bf +}}

\newcommand*\dash{\ifvmode\quitvmode\else\unskip\kern.16667em\fi---%
\hskip.16667em\relax}

\newenvironment{sarahlist}{
\begin{description}[itemsep=2pt,leftmargin=0.4cm]
}{\end{description}}

\newcommand\changestrategy{\mathsf{change}_\cluster}
\newcommand\tx{\mathsf{tx}}

\newcommand\candidates{\mathsf{candidates}}

\newcommand\addr{\mathsf{addr}}
\newcommand\cluster{\mathsf{C}}
\newcommand\validation{\mathsf{validation}}
\newcommand\expansion{\mathsf{expansion}}
\newcommand\heuristic{\mathsf{heur}}
\newcommand\clustertxs{\cluster_\tx}

\newcommand\expandcluster{\mathsf{expansion}_\cluster}

\newcommand\expansfactor{\mathsf{Expsn}}
\newcommand\accuracy{\mathsf{FDR}}
\newcommand\clusteraddrs{\cluster_\addr}
\newcommand\clustertxfeatures{\mathsf{TF}_{\cluster}}
\newcommand\clusteraddrfeatures{\mathsf{AF}_{\cluster}}

\newcommand\txstart{\mathsf{tx}}
\newcommand\txcur{\mathsf{tx}_\mathsf{cur}}
\newcommand\forwardtxs[1]{\mathsf{fwdTxs}_{#1}}
\newcommand\backwardtxs[1]{\mathsf{bkwdTxs}_{#1}}
\newcommand\txnext{\mathsf{next}}
\newcommand\txprev{\mathsf{prev}}

\newcommand\txinput{\mathsf{input}}
\newcommand\txoutput{\mathsf{output}}
\newcommand\outp{\mathsf{outputs}}
\newcommand\inp{\mathsf{inputs}}
\newcommand\val{\mathsf{value}}

\newcommand\backwardscope{\mathsf{bkwdScope}}

\newcommand\txinputPrevHop{\txinput.\txprev}
\newcommand\txoutputNextHop{\txoutput.\txnext}

\newcommand\txinputValue{\txinput.\val}
\newcommand\txoutputValue{\txoutput.\val}

\newcommand\txinputAddr{\txinput.\addr}
\newcommand\txoutputAddr{\txoutput.\addr}

\newcommand\txinputIdxPrev{\txinput.\mathsf{prevIdx}}

\newcommand\fnext{\mathsf{findNext}}
\newcommand\modfnext{\mathsf{findNext2}}
\newcommand\fprev{\mathsf{findPrev}}
\newcommand\fnextResult{\mathsf{nextTx}}
\newcommand\fprevResult{\mathsf{prevTxs}}

\newcommand\followFwd{\mathsf{followFwd}}
\newcommand\followBwd{\mathsf{followBkwd}}

\newcommand\valCluster{\mathsf{Pchain_V(\txstart)}}
\newcommand\valClusterr{\mathsf{Pchain_V(\cluster)}}

\newcommand\featuresTx{\mathsf{features_\tx}}
\newcommand\featuresAddr{\mathsf{features_\addr}}

\newcommand\vall{\mathsf{Val}_\cluster}

\begin{document}

\title{How to Peel a Million: Validating and Expanding Bitcoin Clusters}

\author[1]{\rm George Kappos}
\author[1]{\rm Haaroon Yousaf}
\author[2]{\rm Rainer Stütz}
\author[2]{\rm Sofia Rollet}
\author[3]{\rm Bernhard Haslhofer}
\author[1]{\rm Sarah Meiklejohn}
\affil[1]{University College London and IC3}
\affil[2]{AIT - Austrian Institute of Technology}
\affil[3]{Complexity Science Hub Vienna}

\maketitle

\begin{abstract}

One of the defining features of Bitcoin and the thousands of cryptocurrencies
that have been derived from it is a globally visible transaction ledger.  While
Bitcoin uses pseudonyms as a way to hide the identity of its participants,
a long line of research has demonstrated that Bitcoin is not anonymous.
This has been perhaps best
exemplified by the development of \emph{clustering heuristics}, which have in
turn given rise to the ability to track the flow of bitcoins as they are sent
from one entity to another.

In this paper, we design a new heuristic that is designed to
track a certain type of flow, called a \emph{peel chain}, that represents many
transactions performed by the same entity; in doing this, we implicitly
cluster these transactions and their associated pseudonyms together.  We then
use this heuristic to both validate and expand the results of existing clustering
heuristics.  We also develop a machine learning-based
validation method and, using a ground-truth dataset, evaluate all our
approaches and compare them with the state of the art.  Ultimately, our goal
is to not only enable more powerful tracking techniques but also call
attention to the limits of anonymity in these systems.

\end{abstract}

\section{Introduction}

Since its introduction in 2008, Bitcoin has used a pseudonymous system for transferring coins, with entities forming transactions in which they and their recipient(s) are identified using just a set of \emph{pseudonyms} or \emph{addresses} that have no inherent link to their identity.  It has been demonstrated by now, however, that this use of pseudonyms does not make Bitcoin anonymous.  This has in large part been driven by the development of various
\emph{clustering heuristics} that identify multiple pseudonyms operated by the
same entity~\cite{Reid2013c,FC:RonSha13,Androulaki2013b,Meiklejohn2013,
Spagnuolo2014a,goldfeder2017cookie,Ermilov2018}, with research also showing that de-anonymization is possible at the network layer~\cite{koshy2014bitcoin,biryukov2014bitcoin}. These clustering heuristics use patterns of usage present in the
Bitcoin blockchain as evidence of the shared ownership of the pseudonyms they cluster together;
one heuristic that has been particularly widely adopted is the
so-called \emph{co-spend} heuristic, which says that all addresses used as
input to the same transaction belong to the same entity.  This heuristic has
been so effective that companies such as Chainalysis now use it \dash and heuristics
derived from it\dash to provide Bitcoin tracking as a service to
both law enforcement agencies and financial institutions, such as cryptocurrency
exchanges, looking to comply with anti-money laundering (AML) regulations.
These heuristics can be used not only to cluster together pseudonyms operated
by the same entity but, as a consequence, to track flows of bitcoins as they
are transferred from one entity to another.

This ability to track flows of bitcoins has been used in several high-profile
investigations, such as the indictment of Ross Ulbricht as the operator of the
Silk Road marketplace~\cite{silkroad-tracking}; the blocked movement of
funds paid to the WannaCry ransomware operators~\cite{wannacry-tracking}; the takedown of one of the largest websites hosting child sexual abuse material~\cite{child-tracking}; and
the takedown of several terrorist financing
campaigns~\cite{terrorist-tracking}.  More recently, Roman Sterlingov was
arrested based on allegations that he served as the operator of the Bitcoin
Fog mixing service for ten years~\cite{bitcoinfog-wired}.  Despite the arrest
taking place in April 2021, the allegations were supported by evidence taken
from the Bitcoin blockchain as early as 2011~\cite{bitcoinfog-affidavit}.
This investigation thus makes clear just what is possible when all
transactions are stored in a globally visible and immutable ledger.

In this paper, we extend known heuristics for tracking flows of bitcoins by
formalizing in Section~\ref{sec:chain} the notion of a \emph{peel chain}, which
is a set of linked transactions that are all initiated by the same entity, and presenting
heuristics for identifying peel chains and following them forwards and
backwards.  As compared with previous heuristics, ours are based not on
properties of individual transactions or addresses within transactions, but
rather on general features associated with a cluster formed by the
co-spend heuristic.  In particular, we describe in
Section~\ref{sec:dataCollection} how we assign features to a cluster based on
the transactions and addresses it contains.

We also describe in Section~\ref{sec:dataCollection} our main dataset, which
consists of 120 clusters formed by the co-spend heuristic.  Using data
provided to us by Chainalysis, we know that 60 of these clusters are
\emph{true positives}, meaning all addresses they contain really do belong to
the same entity, and that 60 of them are \emph{false positives}, meaning they
contain one or more \emph{Coinjoin} transactions and thus all addresses do not
actually belong to the same entity.  This access to ground-truth data provides
us with the rare ability to evaluate the accuracy of our heuristics, as well
as ones that were previously proposed.

In particular, we argue how our basic heuristic for identifying and following
peel chains can be used to both \emph{validate} and \emph{expand} traditional
clustering heuristics such as the co-spend heuristic.  In this first usage,
presented in Section~\ref{sec:validation}, we argue that we can use the
ability to identify peel chains to increase our confidence in the results of
the co-spend heuristic.  We also present
in Section~\ref{sec:dataScience} a machine learning approach for validating a cluster as a whole; i.e., for classifying it as either a true positive or
a false positive.  Using our ground-truth dataset, we are able to evaluate
this classifier and show that it achieves an accuracy of 89\%.

In the second usage as an \emph{expansion} heuristic, we demonstrate how the
ability to identify peel chains can be used to expand the results of the
co-spend heuristic.  In particular, following a
peel chain forwards means identifying the \emph{change output} in
each transaction in the chain, so our algorithm includes an implicit
change heuristic.  As compared with previous change
heuristics~\cite{Androulaki2013b,
Meiklejohn2013,goldfeder2017cookie,Ermilov2018}, we demonstrate using our
ground-truth dataset that our change heuristic achieves a false discovery rate
of only 0.02\%; the next-best heuristic achieves a false discovery rate of
12.7\%.
We then apply this expansion heuristic to investigate ransomware
addresses and to track the funds withdrawn
from an exchange account associated with Roman Sterlingov to their deposit into
the Bitcoin Fog mixing service, showing that our heuristic is able to link
these two transactions whereas all previous heuristics would be unable to do
so.

To summarize, we make the following contributions:

\begin{itemize}
\item We provide a heuristic, based on a robust set of features, for both identifying peel chains and following them forwards and backwards.

\item We present a machine learning-based classifier and a validation heuristic, both of which can be used to influence the confidence we can have in the results of the co-spend heuristic.

\item We present a heuristic for expanding co-spend clusters, and evaluate it using a custom-built ground-truth dataset.  Our comparison with previous heuristics shows that ours is significantly more effective and significantly safer.
\end{itemize}

The techniques we develop are directly applicable in cryptocurrency investigations, and thus have the potential to be adopted and used in them.  Above all, however, we hope that our work helps to correct the misperception of Bitcoin~\cite{krombholz16fc,mai20soups,abramova21chi}
as ``anonymous and almost untraceable''~\cite{wsj-hamas} and a way to allow ``people [to] receive digital payments without revealing their identity''~\cite{newyorker-ransomware}.

\section{Related Work}\label{sec:related}

\subsection{Bitcoin clustering}

The ability to cluster together the addresses used as an input to a transaction was first observed in the original Bitcoin whitepaper~\cite{whitepaper}, and has been used in many subsequent works~\cite{Reid2013c,FC:RonSha13,Androulaki2013b,Meiklejohn2013,Spagnuolo2014a}.  Beyond this \emph{co-spend heuristic}, researchers have developed other heuristics for clustering together Bitcoin addresses.  In particular, Meiklejohn et al.~\cite{Meiklejohn2013} and Androulaki et al.~\cite{Androulaki2013b} defined a heuristic for identifying which output in a Bitcoin transaction represented the change being made; this \emph{change heuristic} then said that this output was controlled by the same entity as the input addresses.  This heuristic was later refined by by Goldfeder et al.~\cite{goldfeder2017cookie} and Ermilov et al.~\cite{Ermilov2018}.
There have been many academic studies using these heuristics in order to track crime~\cite{8418627,portnoff2017backpage,paquet2019spams,10.1093/cybsec/tyz003,huang2014botcoin,236358}.
Finally, in concurrent work M{\"o}ser and Narayanan propose a machine learning-based heuristic for identifying change outputs that uses some of the same transaction and address features as in our work~\cite{malte21}.

Beyond proposing new clustering heuristics, a number of studies have attempted to quantify their effectiveness and accuracy. Nick~\cite{Nick2015a} measured the accuracy of different clustering algorithms using a ground-truth dataset consisting of 37,585 user wallets, which was obtained via a vulnerability in the BitcoinJ light client implementation. The results showed that on average more than 69\% of the addresses could be linked using only the co-spend heuristic. Harrigan and Fretter~\cite{Harrigan2017b} studied reasons for the effectiveness of the co-spend heuristic and concluded that address reuse and avoidable merging were the main drivers.
Fröwis et al.~\cite{Frowis2019e} discussed the effectiveness of the combined use of clustering heuristics and attribution tags as forensic tools.  By empirically quantifying the effect of Coinjoin transactions, they showed that clustering heuristics can lead to false interpretation and pointed to the need for additional metrics to quantify the reliability of clustering results.

\subsection{Bitcoin entity classification}
Bartoletti et al. \cite{Bartoletti2018} investigated data mining techniques to automatically detect Ponzi schemes carried out using Bitcoin.  Using supervised learning algorithms, the authors could correctly classify Ponzi schemes with a very low rate of false positives.  A number of other studies use supervised learning in order to classify unknown addresses according to the entities they belong to~\cite{8258365,harlev2018breaking, YinRegulating}.
Unsupervised learning methods have also been used in Bitcoin to attempt to identify fraud~\cite{Pham2014,Pham2016,Zambre2013}.

Ranshous et al.~\cite{Ranshous2017b} applied the notion of motifs in directed hypergraphs to identify distinct statistical properties related to Bitcoin exchange addresses. They build different classification models (Random Forest, AdaBoost, Linear SVM, Perceptron, Logistic Regression) using a set of features, obtaining the best results with Random Forest and AdaBoost.
Jourdan et al.~\cite{Jourdan2018} consider the more general problem of classifying entities in multiple classes on the basis of properties of their extended transaction neighborhoods.

\section{Background}\label{sec:back}

\subsection{Bitcoin transactions}\label{sec:fundamentals}

A Bitcoin transaction $\tx$ consists of ordered lists of inputs ($\tx.\inp$) and outputs ($\tx.\outp$). We use $\tx.\inp[i]$ to denote the $i$-th input and do the same for $\tx.\outp$.  Each output $\txoutput$ in a transaction is associated with an address $\txoutputAddr$ and a value $\txoutputValue$ representing the amount of coins received by the address in this transaction. Each input to a transaction is similarly associated with an address $\txinputAddr$ and a value $\txinputValue$ representing the amount of coins being sent by the address in this transaction.  Furthermore, an input is itself a \emph{transaction output} (\emph{TXO}); i.e. an input points to the transaction in which its associated address received coins.  Other peers in Bitcoin's peer-to-peer network can then check that all inputs to a transaction are well formed and are \emph{unspent} (meaning they are \emph{UTXOs}), before propagating the transaction to other peers and eventually enabling its inclusion in the blockchain.  This ensures that double-spending is not included in the blockchain, which acts as a global ledger of all transactions.

Concretely, this means that transactions point both \emph{backwards}, in terms of the input UTXOs pointing to the past transactions in which they received the coins they are now spending, and \emph{forwards}, in terms of the UTXOs in future transactions that reference any output addresses that get spent.  We denote the transaction that an input $\txinput$ points backwards to by $\txinputPrevHop$, and the transaction that an output $\txoutput$ points forwards to by $\txoutputNextHop$ (if it is spent).  We also denote by $\txinputIdxPrev$ the index of an input $\txinput$ in $\txinputPrevHop.\outp$; i.e., the index of its transaction output in the transaction in which it was created.  For the rest of this work, when we refer to \emph{following} an input to a transaction we mean looking backwards at the past transaction that created this UTXO, and when we refer to following an output we mean looking forwards to any UTXOs that reference it.

\subsection{Clustering Bitcoin addresses}\label{sec:clustergraphs}

A valid Bitcoin transaction needs to be signed using the private keys associated with all its inputs.  This has given rise to a common heuristic for clustering together Bitcoin addresses, known as the multi-input or \emph{co-spend} heuristic~\cite{Reid2013c,FC:RonSha13,Androulaki2013b,Meiklejohn2013,Spagnuolo2014a}.  This heuristic states that all inputs to a transaction are controlled by the same entity, using the fact that they have all signed the transaction as evidence of shared ownership.

Although this heuristic is considered safe in general and has been adopted in practice, it can be invalidated by a specific type of transaction called a \emph{Coinjoin}.  When forming a Coinjoin, users work together to create a transaction in which they each control a different input and the outputs likewise represent different recipients.  This acts to mix together the coins of these users and thus destroys the link between each individual sender and recipient.  Furthermore, it invalidates the co-spend heuristic as it is no longer the case that all inputs are controlled by the same entity.

Beyond the co-spend heuristic, there are a number of proposed \emph{change} heuristics~\cite{Meiklejohn2013,Androulaki2013b,Ermilov2018}; i.e., heuristics for identifying which output in a transaction that the sender uses to send themselves their change (the value of their UTXO subtracted by the amount they are sending to the recipient).  As observed by Meiklejohn et al.~\cite{Meiklejohn2013}, identifying such outputs not only makes it possible to include this address in the same cluster as the sender and thus enhance the co-spend heuristic, it also makes it possible to identify that the transaction in which this change output is spent is also performed by the same entity.  We describe this pattern of following \emph{peel chains} in more detail in Section~\ref{sec:chain}, and describe these proposed change heuristics in more detail in Section~\ref{sec:expansion-results} when we compare our own heuristic against them.

\section{Dataset and Methodology}\label{sec:dataCollection}

To start, we were given 241 Bitcoin addresses and 20,016 Bitcoin transactions by Chainalysis, a company that provides blockchain
data and analysis to businesses and government agencies.\footnote{\url{https://www.chainalysis.com/}}
The addresses represented \emph{true positive} clusters, in the sense that Chainalysis had manually verified that all the addresses in the same co-spend cluster as this address really did belong to the same service (typically by confirming directly with the service).  The transactions were all Coinjoins and thus represented \emph{false positive} clusters, meaning all of the addresses in the resulting co-spend cluster would not actually belong to the same service.  Each address formed a distinct cluster, and there was no overlap between the addresses in the true positive (TP) clusters and the ones used as inputs in the false positive (FP) transactions.  This ground-truth dataset was necessary for evaluating our heuristics, and would not have been possible to get at this scale without working with Chainalysis or directly with the services themselves. None of the clusters represented individual users, and we had no additional information about the entities represented by the clusters (e.g., the name of the service).

From this initial dataset, we created clusters using the co-spend heuristic and represented each cluster $\cluster$ as a tuple $(\clusteraddrs, \clustertxs)$ where $\clusteraddrs$ is the set of all addresses in the cluster and $\clustertxs$ is the set of all transactions initiated by one or more addresses in $\clusteraddrs$.  This resulted in 241 TP clusters and 16,974 false positive FP clusters.
We describe in Section~\ref{sec:tpfp} how from this initial dataset we created a more balanced dataset of 60~true positive (TP) and 60 false positive (FP) clusters that we then used in the remainder of our analysis.  In order to do so, we first describe the features we defined for transactions, addresses, and the overall cluster.

\subsection{Features}

We consider features of three different types of objects within Bitcoin: transactions, addresses, and clusters.  These features are largely defined by the wallet software used by a given entity and the decisions they make in scripting their transactions, and as we will see the set of possible features is largely stable within even large clusters.  This consistency is crucial in the algorithms we develop for following \emph{peel chains} in Section~\ref{sec:chain}.  Our set of chosen features is based on Bitcoin usage today, but we stress that new features can be incorporated as Bitcoin evolves without changing our overall approach.

\subsubsection{Transaction features}\label{sec:txFeatures}

Every Bitcoin transaction has a different set of features, according to
both the action it is performing and the wallet program and version used to
generate it.  We consider the following four features.

\begin{sarahlist}
\item[Replace-by-fee/sequence number.] At any given point in time,
there can be multiple versions of the same transaction in the Bitcoin network;
for example, if a user broadcasts a transaction to the network but it never
gets included in a block, they may broadcast a new version with an increased
fee in the hopes of increasing its chances. The \emph{sequence number} helps
identify different versions of a transaction, with a higher sequence number
indicating that the transaction is more recent. If a user does not want
transactions to be able to be replaced they can thus set the sequence number
to be the maximum value (\verb#0xffffffff#).  The sequence number is set for each transaction input, and the transaction is considered to be \emph{replaceable} if any of its inputs have a sequence number less than this maximum value~\cite{bip0125}.  We thus set this feature to be true for a transaction if it is replaceable and false if it is not.

\item[Locktime.] A transaction can set a \emph{locktime} (or time lock) to indicate that it cannot be spent before a block at some height has been mined.  We set this feature to be true if a locktime has been set and false if not.

\item[Version.] The \emph{version} of a transaction, which is either 1 or 2, determines the rules used to validate the transaction~\cite{bip68}.

\item[SegWit.] \emph{SegWit} (Segregated Witness)~\cite{bip141} allows a transaction to be separated into its semantic data (i.e.,
information about who is sending and receiving bitcoins) and its signature
data.
A transaction can indicate if it uses SegWit by setting its fifth byte to \verb#0x00#.  We set this feature to be true if SegWit is enabled and false if not.

\end{sarahlist}

We thus represent the features of a transaction $\tx$ as a 4-tuple containing binary values (1/2 for the version and true/false for the rest) in each entry.
We denote by $\featuresTx$ the function used to extract these features from a transaction.

\subsubsection{Address features}\label{sec:addrFeats}

The BlockSci tool~\cite{blocksci} categorizes Bitcoin scripts into ten generic types:
\textsf{pubkey}, \textsf{pubkey hash}, \textsf{witness pubkey hash},
\textsf{multisig}, \textsf{multisig pubkey}, \textsf{script hash}, \textsf{witness script hash}, \textsf{witness unknown},
\textsf{non-standard}, and \textsf{nulldata} (with the \textsf{witness} prefix indicating that it uses SegWit).
Some of these categories can be further broken down according to whether the address is \emph{compressed} or \emph{uncompressed}.
To briefly explain some of the more common types, the \textsf{pubkey} format allows users to send coins to a public key.  Both \textsf{pubkey hash} and \textsf{witness pubkey hash} allow users to instead send coins to the hash of a public key.
The \textsf{script hash} and \textsf{witness script hash} formats allow users to send coins to the hash of an arbitrary script.  These coins can then be spent only by the owner(s) of the underlying script.  A common script in Bitcoin is an $m$-of-$n$ \textsf{multisig}.  Using a \textsf{multisig} address, a user or set of users can require that at least $m$ of the $n$ available keys specified in the script must sign a transaction in order to spend the coins from that address.

Across our set of 246,600 addresses, we identified 10 distinct combinations of these categories that were used, as summarized in Table~\ref{tab:address-count}.  We thus represent the features of an address as its address type, which takes one of these ten values.
We denote by $\featuresAddr$ the function used to extract the feature from an address.

\begin{table}[t]
  \centering%
\begin{tabular}{lS[table-format=2.2]S[table-format=2.2]}
  \toprule
Address type & {TP (\%)} & {FP (\%)} \\
  \midrule
\textsf{pubkey hash} (compressed)            & 41.15 & 1.19 \\
\textsf{pubkey hash} (uncompressed)          &   0.0  & 0.010 \\
\textsf{witness pubkey hash} (compressed)    &  5.76 & 37.44 \\
\textsf{witness pubkey hash} (uncompressed)  &  37.04 & 61.36 \\
\textsf{multisig} (2/2) &  5.6 & 0.0 \\
\textsf{multisig} (2/3) &  2.8 & 0.0 \\
\textsf{multisig} (3/4) &  0.1 & 0.0 \\
\textsf{multisig} (2/6) &  0.24 & 0.0 \\
SegWit \textsf{multisig} (2/2)  &  2.22 & 0.0 \\
SegWit \textsf{multisig} (2/3)  &  5.12 & 0.0 \\
  \bottomrule
\end{tabular}
  \caption{Types of addresses found across all clusters.}%
  \label{tab:address-count}%
\end{table}

\subsubsection{Cluster features}\label{sec:clusFeats}

Each cluster contains a set of addresses and a set of transactions.  From these sets $\clustertxs$ and $\clusteraddrs$, we can extract the relevant features (which we did using BlockSci) to build the sets $\clustertxfeatures$ and $\clusteraddrfeatures$ of all transaction and address features present in the cluster.

In addition to these sets of features, we define for each cluster a \emph{change strategy}, which we denote by $\changestrategy$.  This cluster-level feature considers the pattern, if any, the transactions in this cluster exhibit when forming change outputs.
We identify a change output in a transaction in $\clustertxs$ if there is exactly one output whose address belongs to the cluster (i.e., is in $\clusteraddrs$).  If there are zero or multiple such addresses then we ignore this transaction for the purposes of setting the change strategy.

Intuitively, some wallet software may send the change in a transaction to a
specific output index by default; e.g., the first or last output. As many
entities are likely to use scripts or other automated methods to
form transactions, it may be the case that their transactions thus have patterns in terms of the index of the change
output.
To this end, we define four different values for the cluster's $\changestrategy$:

\begin{sarahlist}
\item[$\changestrategy = -1$.] For every transaction in $\clustertxs$ with a single identified change output, it was always at the last index.

\item[$\changestrategy = 0$.] For every transaction in $\clustertxs$ with a single identified change output, it was always at the first index.

\item[$\changestrategy = 1$.] For every transaction in $\clustertxs$ with a single identified change output, it was always at either the first or the last index.

\item[$\changestrategy = {\normalfont \text{None}}$.] There was at least one transaction in
$\clustertxs$ that did not follow the patterns above; i.e., with a single identified change output that was at neither the first nor the last index.
\end{sarahlist}

\subsection{Creating a cluster dataset}\label{sec:tpfp}

In building a dataset of clusters, our goal was to have it be as balanced as
possible, in terms of the features introduced in the previous section.  This
was particularly important in our validation of the co-spend heuristic in
Section~\ref{sec:overall-validation}, in which we differentiate between TP and
FP clusters based on their features and need to avoid overfitting.  Concretely, we focused on creating a balance between true and false positives for the following three parameters: (1) the number of clusters in each category, (2) the sizes of both $\clusteraddrs$ and $\clustertxs$, and (3) the period of time in which which the cluster was active.  We refer to the last property as the cluster's \emph{lifespan}.
The first two properties are generally important in creating a balanced dataset, and
this last property is also essential as behavior in Bitcoin transactions has changed significantly over time.
Ensuring that the lifespans of TP and FP clusters had a significant overlap was thus the only way to ensure a fair comparison; e.g., making it so we could not trivially distinguish because all TP clusters had one transaction feature set to true and all FP clusters had it set to false.

\begin{figure*}[t!]
\centering
\begin{subfigure}[b]{0.21\textwidth}
   \includegraphics[width=\linewidth,trim=0 -86 0 0,clip]{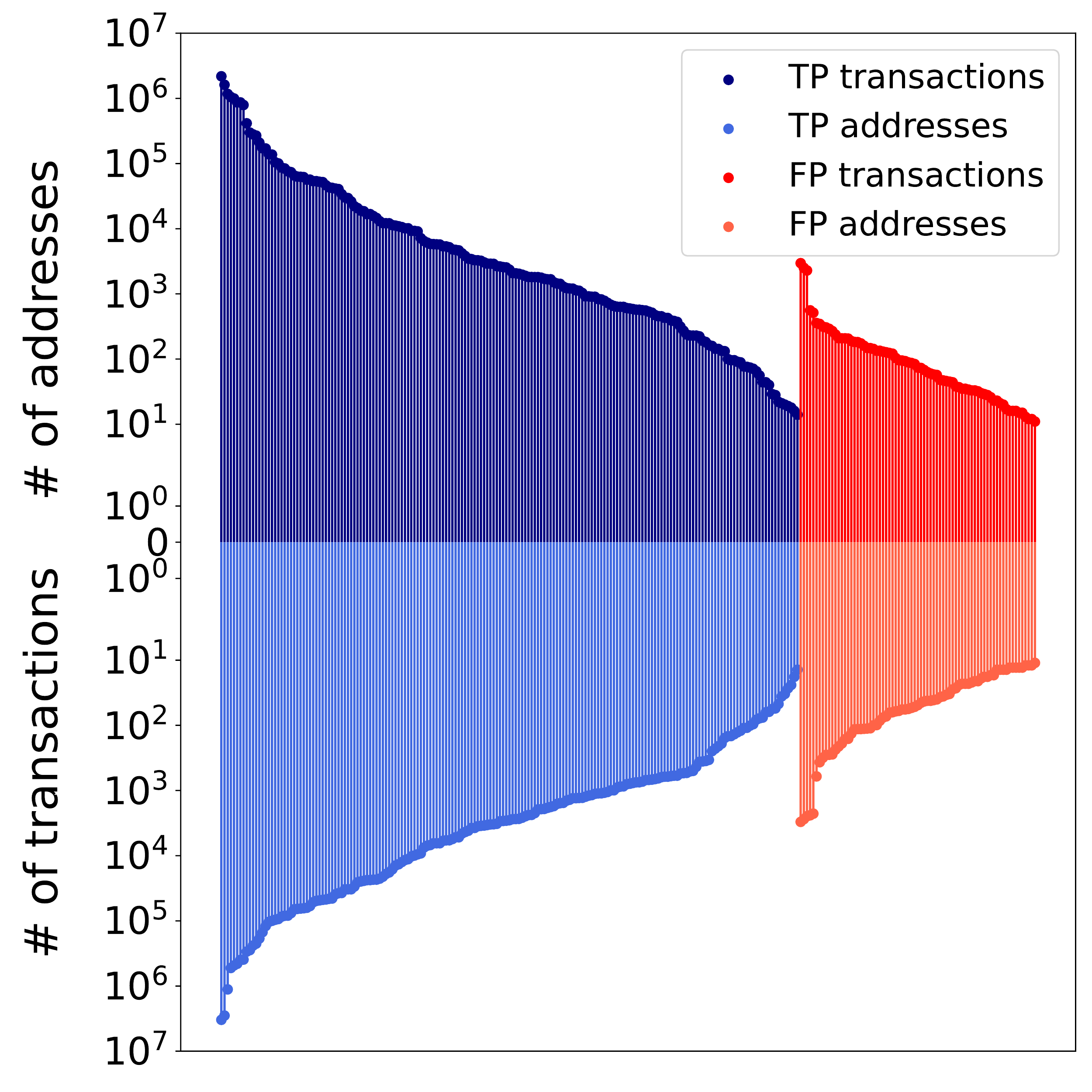}
    \caption{}
   \label{fig:Ng1}
\end{subfigure}
\hspace{5mm}
\begin{subfigure}[b]{0.21\textwidth}
   \includegraphics[width=\linewidth,trim=0 -86 0 0,clip]{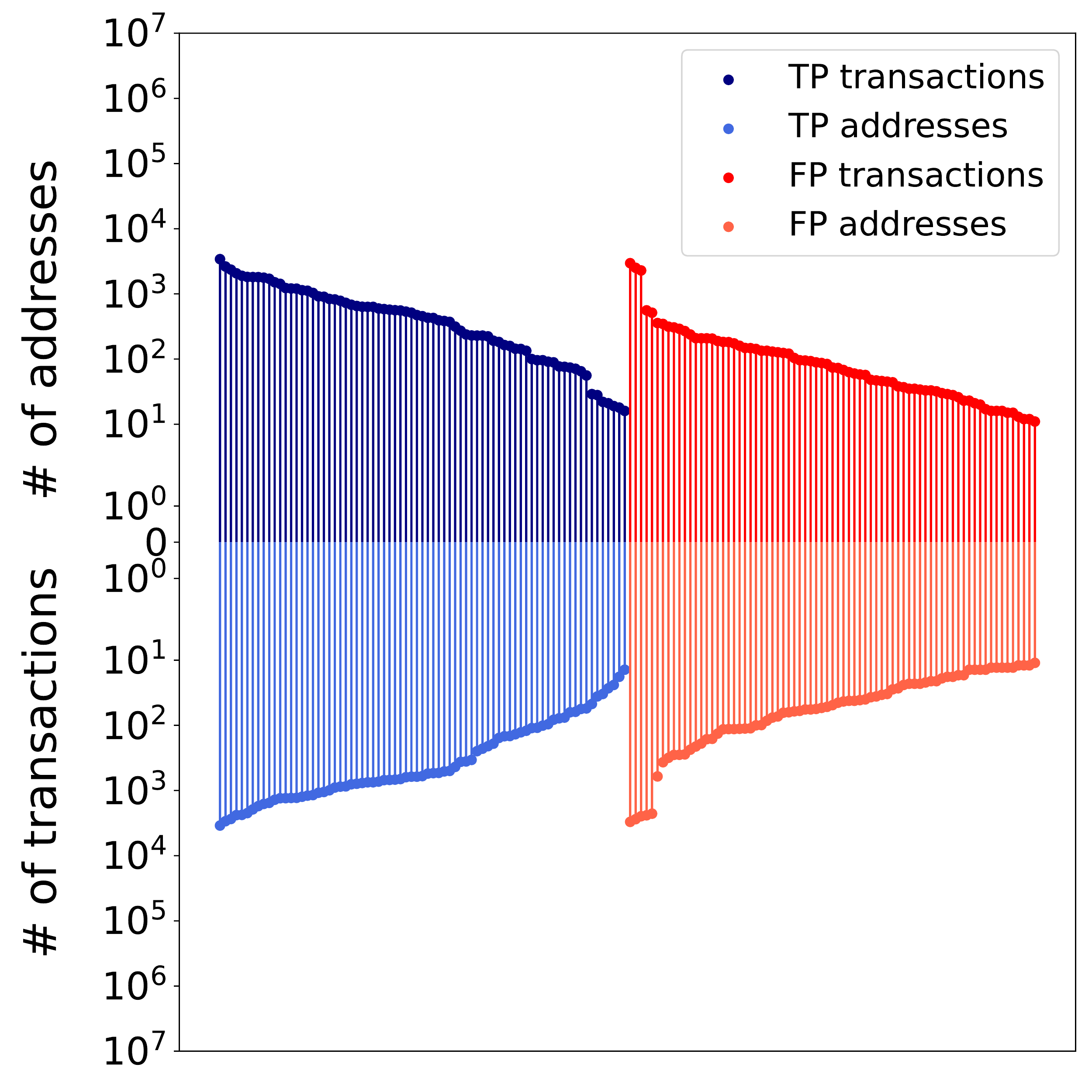}
    \caption{}
   \label{fig:Ng2}
\end{subfigure}
\begin{subfigure}[b]{0.26\textwidth}
   \includegraphics[width=\linewidth]{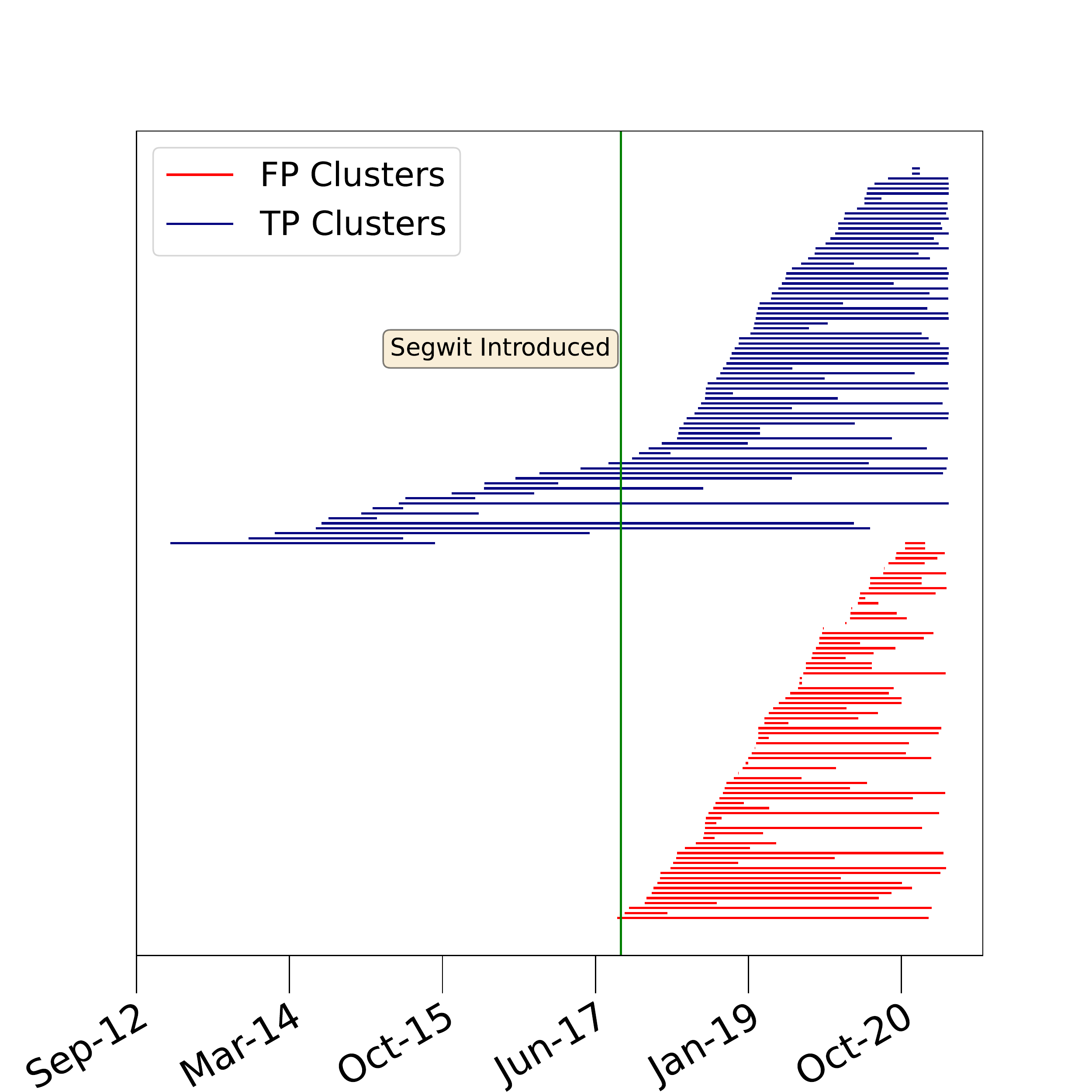}
   \caption{}
   \label{fig:Ng3}
\end{subfigure}
\begin{subfigure}[b]{0.26\textwidth}
   \includegraphics[width=\linewidth]{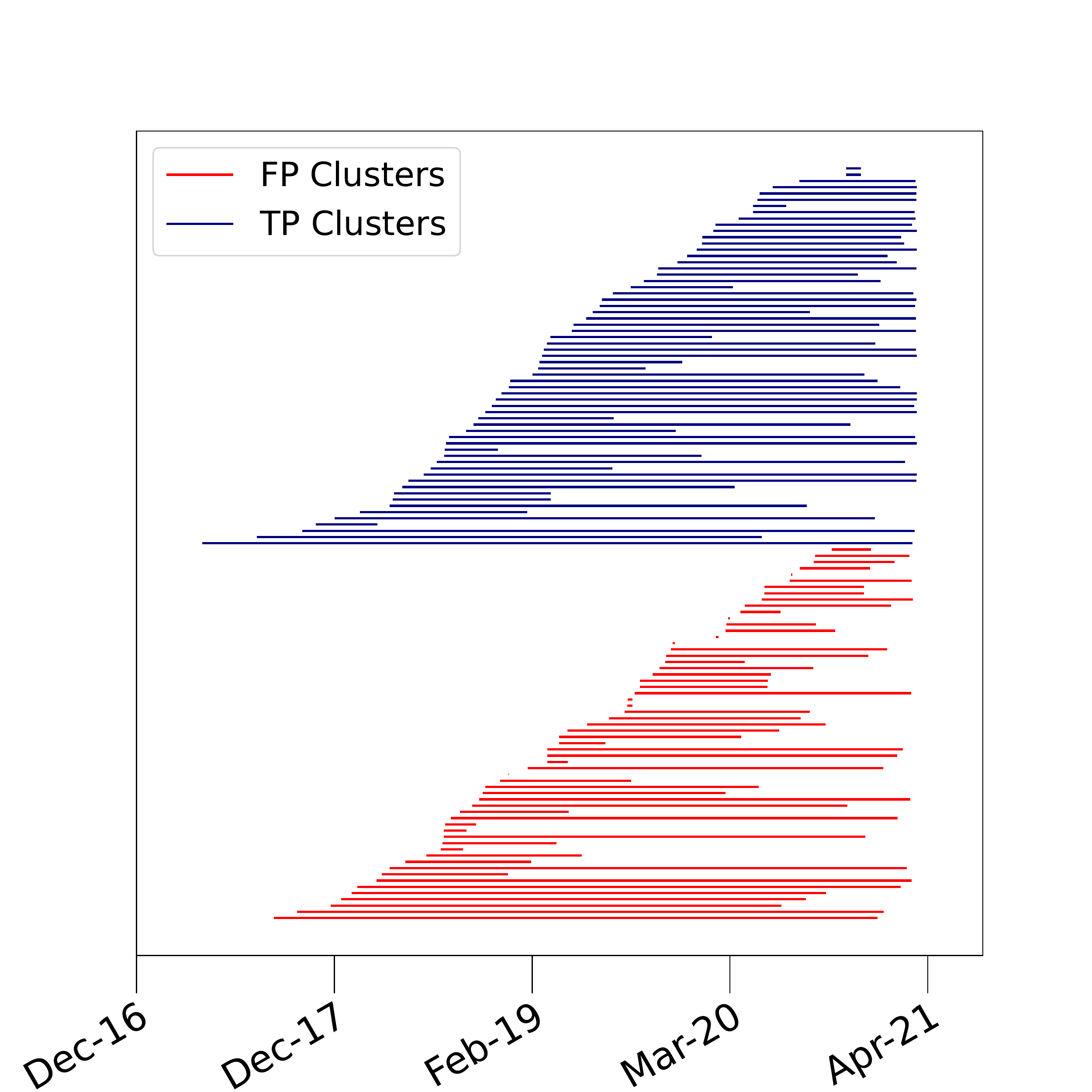}
   \caption{}
   \label{fig:Ng4}
\end{subfigure}
\caption{Balancing the sizes (Figures~\ref{fig:Ng1} and~\ref{fig:Ng2}, with the $y$-axis on a log scale) and lifespans (Figures~\ref{fig:Ng3} and~\ref{fig:Ng4}) for our true positive (TP) and false positive (FP) clusters.  Figures~\ref{fig:Ng1} and~\ref{fig:Ng3} represent the cluster sizes and lifespans before balancing, and Figures~\ref{fig:Ng3} and~\ref{fig:Ng4} represent these values after balancing.  In Figure~\ref{fig:Ng3}, the vertical line represents the date on which SegWit was introduced.}
\end{figure*}

To start, we set a minimum threshold of 10 for both $\clustertxs$ and $\clusteraddrs$ and discarded all clusters that were smaller.  This left us with 183 TP clusters (out of 241) but only 75 FP clusters (out of 16,974).  This overrepresentation of singleton FP clusters was largely due to the way in which we obtained data from Chainalysis, as we asked for false positive transactions (i.e., Coinjoins) but true positive addresses that would form a meaningful cluster (i.e., not a singleton).  Additionally, the vast majority of the inputs to the Coinjoin transactions were one-time addresses, meaning the resulting cluster was such that $|\clustertxs|=1$.
After obtaining these 183 TP and 75 FP clusters, we further observed that they were highly imbalanced in the parameters we considered, as shown in \Cref{fig:Ng1}.

This figure shows that not only are there more TP clusters, but also they are much larger on average than the FP clusters.  For example, in our initial dataset, TP clusters had up to 3.2M transactions (with an average of 56K) whereas FP clusters had only up to 283K (with an average of 19K).
To this end, we removed the 108 biggest TP clusters from our analysis
(in terms of $|\clusteraddrs| + |\clustertxs|$).  This created a dataset of 75 TP and 75 FP clusters, each of comparable size (in terms of both $\clusteraddrs$ and $\clustertxs$), as we see in \Cref{fig:Ng2}.

This approach created a balance in terms of the first property, but did not address the issue of having overlapping lifespans: as we see in \Cref{fig:Ng3}, there are several TP clusters whose lifespan ended before any FP clusters even began.  We can also see this has a direct impact on our transaction features, as several of our TP clusters existed before SegWit was introduced but none of our FP clusters did.
To address this imbalance, we
removed the TP clusters whose lifespan didn't overlap
with the lifespan of any FP clusters. We ended up with 60
TP and 60 FP clusters that were balanced in all three properties, as shown in \Cref{fig:Ng2} and \Cref{fig:Ng4}.  In the end, all clusters had between 15 and 3415 addresses (with an average of 258.6 for FP clusters and 642.6 for TP ones), between 11 and 3448 transactions (with an average of 307.7 for FP clusters and 692.4 for TP ones), and operated at some point between April 2017 and April 2021.

\paragraph{Feature statistics.}
In terms of transaction features, we found each of the possible 16 4-tuples in at least one of our clusters.  Most clusters (76 out of 120) used only a single combination of transaction features, however, and all clusters used six or fewer.  The average number of features was 1.55 for FP clusters and 1.67 for TP clusters.
This suggests that clusters are largely consistent in their transaction behavior.
We found a similar level of consistency when looking at address features: 56.7\% of TP clusters and 53.3\% of FP clusters used only one address type, and all clusters used three or fewer.

We found that $18$ of our TP clusters had a completely consistent change strategy ($\changestrategy=0$ or $-1$) and $30$ had $\changestrategy=1$; this left $12$ with no change strategy.  As might be expected, FP clusters were less consistent (given that they actually contained different sets of users): 34 had no identifiable change strategy, 8 had a completely consistent change strategy, and 18 had $\changestrategy=1$.

To understand the overlap in features across different clusters, we looked at the Jaccard similarity between the sets of 5-tuples representing their combined transaction and address features.  We found that the average Jaccard similarity across all pairs of clusters was $0.13$, which suggests that they are relatively dissimilar in these features.  There were several notable exceptions, however, and in particular cases where the features were not only overlapping but in fact identical.  For example, there were 26 clusters whose transactions all had the same combination of features (Version~1 and SegWit-enabled transactions that had no locktime and were not replaceable) and whose addresses were all of the same type (uncompressed \textsf{witness pubkey hash}).  This is unsurprising as it represents the default setting of the standard Bitcoin wallet software.

\section{Following Peel Chains}\label{sec:chain}

In this section, we define the concept of a peel chain and present our heuristics for identifying them, according to the features defined in the previous section.

\subsection{Defining a peel chain}

The concept of a peel chain was first introduced by Meiklejohn et al.~\cite{Meiklejohn2013} as a series of transactions originating from a transaction with a relatively large UTXO as input (i.e., a UTXO with a large associated value).  When this UTXO is spent it creates two outputs: a small one representing a payment to an external entity and a larger one representing the change.  This pattern can then be repeated many times, with each hop in the chain slowly ``peeling'' smaller values from the original UTXO, until the remaining change amount is small (at which point it can be combined with other small UTXOs to create a large one and start the process over again).
In this work we consider more general peel chains in which transactions might have more than one input and more than two outputs.  In fact, our only requirement is that each adjacent hop in the peel chain is connected by a change output.

Peel chains are in some sense fundamental to UTXO-based cryptocurrencies, which do not allow the partial spending of transaction outputs.  Given that it is highly unlikely for an entity to have the exact amount they want to pay someone associated with a UTXO, their payment inevitably forms a change output that can in turn be used as input to a subsequent payment.
A second reason that peel chains are very common in Bitcoin is more specific to bigger services, as we explore in \Cref{sec:validation}. In particular, it would be time-consuming, error-prone, and inefficient for big services to craft thousands of transactions manually. For these reasons, they naturally perform transactions using scripts.  This automated behavior not only creates long peel chains but also creates patterns that make these peel chains easier to identify and follow.

\subsection{Identifying inputs and outputs}\label{sec:one-hop}

In order to follow peel chains, the first step is to link together
transactions according to their change outputs.  Concretely, this means
introducing two algorithms: $\fnext$, which aims to
identify the unique change output in a transaction, and $\fprev$,
which aims to identify the input(s) in a transaction that originate from
transactions conducted by the same entity (as opposed to transactions in which
that entity received coins from another one).  In its goal, $\fnext$ is
comparable to previous work that developed change identification
heuristics~\cite{Androulaki2013b, Meiklejohn2013, goldfeder2017cookie, Ermilov2018}.  As we describe in more detail in Section~\ref{sec:expansion-results}, however, these
previous works identify change
outputs based on the freshness of output addresses and the output values.  In contrast, $\fnext$ focuses on the index of the output (according to $\changestrategy$), the
cluster's features (according to $\clustertxfeatures$ and
$\clusteraddrfeatures$), and the next hops of every output that has been
spent.

We formally specify $\fnext$ and $\fprev$ in Algorithms~\ref{algo:fnext}
and~\ref{algo:fprev}.  Intuitively, both start with a transaction
$\tx\in\clustertxs$, and aim to output either the next hop in this
transaction's peel chain ($\fnext$) or the previous hops ($\fprev$), according to the features exhibited by the cluster.

\paragraph{$\fnext$.}
For $\fnext$, we first identify the set of outputs that
might represent the change output, according to the cluster's change strategy
$\changestrategy$ (lines~1--6 in Algorithm~\ref{algo:fprev}).  If the change
strategy is associated with a single output index~$i$ (either $0$ or $-1$)
then we include only that output in this candidate set, while if it $1$ we
include both the first and last outputs and if it is $\text{None}$ we include
all outputs. Next, for each candidate change output $\txoutput$ we check to
see if:

\begin{enumerate}

\item The output is spent, meaning $\txoutputNextHop \neq \bot$.

\item The address has a type that exists within the
cluster, meaning $\featuresAddr(\txoutputAddr) \in \clusteraddrfeatures$.

\item The next hop of the output has features
that exist within the cluster, meaning $\featuresTx(\txoutputNextHop) \in
\clustertxfeatures$.

\end{enumerate}

If each of these checks pass, we add the transaction in which this output
is spent to a set representing the possible next hops in the
peel chain (line~13). At the end, if there is only one candidate transaction
then we output it as the next hop.  If there are zero or multiple choices
then we output $\bot$ to indicate that we are unsure of the change output.

\begin{algorithm}[t!]
\SetAlgoLined
\KwResult{ $\fnextResult$ }
\SetKwInOut{Input}{Input}
\Input{ $\tx$, $\changestrategy$, $\clustertxfeatures$, $\clusteraddrfeatures$ }
\uIf{$\changestrategy\in\{0,-1\}$}{
    $\candidates \leftarrow \{\tx.\outp[\changestrategy]\}$
}\uElseIf{$\changestrategy$ = $1$}{
    $\candidates \leftarrow \{\tx.\outp[0], \tx.\outp[-1]\}$
}
\Else{
    $\candidates \leftarrow \tx.\outp$
}
$\fnextResult\leftarrow \emptyset$

\For{$\txoutput \in \candidates$}{
    $b_\txnext\gets (\txoutputNextHop \neq \bot)$

    $b_\addr\gets (\featuresAddr(\txoutput.\addr) \in \clusteraddrfeatures)$

    $b_\tx\gets(\featuresTx(\txoutputNextHop) \in \clustertxfeatures)$

    \uIf{$b_\addr \land b_\txnext \land b_\tx$} {
    $\fnextResult \gets \fnextResult \cup \{\txoutputNextHop\}$
    }
}
\uIf{$|\fnextResult| = 1$}{
    \Return $\fnextResult[0]$
}
\Else{
    \Return $\bot$
    }
\caption{$\fnext$}
 \label{algo:fnext}
\end{algorithm}

We experimentally evaluate the accuracy of $\fnext$ in Section~\ref{sec:expansion-results}, where we see it produces a very low number of false positives.  To see why, we consider that $\fnext$ incorrectly identifies the next hop in a peel chain only if two things happen simultaneously: (1) the transaction either doesn't produce a change output or the cluster deviates from its known address and transaction features only in spending the change output in $\tx$, and (2) exactly one output that meaningfully receives coins in $\tx$ has the same address features as $\cluster$, and produces the same transaction features when it spends the received coins.  In other words, an entity would have to change its established behavior at the same time as it sends coins to another entity with the exact same features.  For the first point, as we discussed above it is unlikely for a transaction to have no change output given that one bitcoin is highly divisible (to the eighth decimal place) and an entity would have to have the exact amount (plus fees) that they wanted to pay someone associated with a UTXO.  As we saw in Section~\ref{sec:tpfp}, clusters are highly consistent in both their address and transaction features, which also makes deviations in their behavior unlikely.  For the second point, we also saw in Section~\ref{sec:tpfp} that clusters are largely non-overlapping in their behavior (with some exceptions).

In terms of false negatives, $\fnext$ fails to identify the change output if either (1) it finds no suitable candidate or
(2) it finds more than one candidate.
In the first case, the cluster would need to either use a different change strategy $\changestrategy$ (putting the change output at a different index from expected) or use a different set of features in both the address and the next transaction.  In the second case,
there needs to be at least one receiving output that behaves in the same way
as $\cluster$, in terms of having the same address and transaction features.  It also needs to be the case that $\cluster$ has $\changestrategy = 1$ or $\changestrategy = \text{None}$, because in the
case where $\changestrategy = 0$ or $\changestrategy = -1$, there is no chance of
$\fnext$ finding multiple candidates since only one will be investigated.  As with false positives, the consistent and distinct qualities of cluster features thus suggest that false negatives are relatively unlikely to occur as well.

\paragraph{$\fprev$.}

Our second algorithm, $\fprev$, looks at the inputs to a transaction rather
than at its outputs.  In particular, while the co-spend heuristic tells us
that each input belongs to the same entity, it may be the case that some of
these inputs represent coins received from other entities.  Our goal is to be
able to follow peel chains (which are created by a single entity) backwards,
which means $\fprev$ must thus isolate the previous transactions in which these
inputs were used as change outputs.  This means that we first map each input
to a transaction $\tx$ to the transaction in which it was created
($\txinputPrevHop$) and to its index in the output list of that transaction
($\txinputIdxPrev$).  We next filter out all previous transactions that do
not match the transaction features of the cluster (line~3 in
Algorithm~\ref{algo:fprev}), and then within this filtered set keep track of
all transactions ($\candidates$), in addition to all transactions in which one
of the inputs was created at either the first or last index ($\candidates_0$
and $\candidates_{-1}$ respectively).  Then, as with $\fnext$, we consider the
change strategy defined by the cluster and use it to decide which of these
candidate sets to return (lines~8--13).

\begin{algorithm}[t!]
\SetAlgoLined
\KwResult{ $\fprevResult$ }
\SetKwInOut{Input}{Input}
\Input{ $\tx$, $\changestrategy$, $\clustertxfeatures$, $\clusteraddrfeatures$ }
$\candidates_0, \candidates_{-1}, \candidates\leftarrow \emptyset$

\For{$\txinput\in\tx.\inp$}{
\uIf{$\featuresTx(\txinputPrevHop) \in \clustertxfeatures$}{
    $i \gets \txinputIdxPrev$

    \uIf{$i \in\{0, -1\}$} {
        $\candidates_i \gets \candidates_i \cup
        \{\txinputPrevHop\}$
    }
    $\candidates\gets \candidates\cup \{\txinputPrevHop\}$
}
}
\uIf{$\changestrategy \in\{0, -1\}$} {
    \Return $\candidates_{\changestrategy}$
}
\uElseIf{$\changestrategy = 1$} {
    \Return $\candidates_0\cup\candidates_{-1}$
}
\uElse{
    \Return $\candidates$
}
\caption{$\fprev$}
 \label{algo:fprev}
\end{algorithm}

The potential for false positives in $\fprev$ is significantly higher than for $\fnext$, as the algorithm returns multiple transactions rather than a single one.  Thus, false positives can occur if any of the entities sending coins in a previous hop exhibits the same transaction features and follows the same change strategy.  In other words, we rely more heavily on cluster features being distinct (as compared to $\fnext$ where we also could count on their consistency), which as we saw in Section~\ref{sec:tpfp} is not always the case.  We discuss this further in Section~\ref{sec:tracingHeuristic}.

In terms of false negatives, $\fprev$ fails to identify a $\txinputPrevHop$ originating from the same cluster only if that $\txinputPrevHop$ deviates in its transaction features or follows a different $\changestrategy$.  Here again we can rely on the consistency of cluster transaction features to argue that this is relatively unlikely to happen.

\subsection{Following transactions}

With $\fnext$ and $\fprev$ in place, we can define algorithms for following peel chains forwards ($\followFwd$) and backwards ($\followBwd$). The ability to follow a transaction both forwards and backwards allows us to capture the full peel chain, regardless of the position of our starting transaction. These algorithms are defined in Algorithms~\ref{algo:followF} and~\ref{algo:followB}.

\paragraph{$\followFwd$.}
To follow a transaction $\txstart$ forwards, $\followFwd$ continues going to the next hop in the peel chain, as identified by $\fnext$, until the peel chain ends or $\fnext$ otherwise cannot identify a next hop. Along the way it adds the hops to a set of transactions $\forwardtxs{\txstart}$, which it outputs at the end.
Line~4 of this algorithm includes a check that is specific to our validation heuristic; we describe this modification in Section~\ref{sec:validation} when we present that heuristic.

\begin{algorithm}[t!]
\SetAlgoLined
\KwResult{$\forwardtxs{\txstart,\heuristic}$}
\SetKwInOut{Input}{Input}
\Input{ $\txstart$, $\heuristic$, $\clustertxs$,
$\changestrategy$, $\clustertxfeatures$, $\clusteraddrfeatures$}

$\forwardtxs{\txstart,\heuristic} \leftarrow \emptyset$

$\txcur \gets \txstart$

\While{$\txcur\neq \bot$} {
    \uIf{$\heuristic = \mathsf{validation}~\land~\txcur\notin \clustertxs$}{
        \Break}

    $\forwardtxs{\txstart,\heuristic} \gets \forwardtxs{\txstart,\heuristic} \cup \{\txcur\}$

   $\txcur \leftarrow \fnext(\txcur,
\changestrategy, \clustertxfeatures, \clusteraddrfeatures)$

    }

\Return $\forwardtxs{\txstart,\heuristic}$
\caption{$\followFwd$}
 \label{algo:followF}
\end{algorithm}

\paragraph{$\followBwd$.}
Following transactions backwards is more involved than following them forwards, as $\fprev$ outputs a set of transactions rather than a single one.  We can think of $\followBwd$ as performing a breadth-first search: it defines a set of transactions to follow, which is initially set to be just the starting transaction (line~2 of Algorithm~\ref{algo:followB}). As long as there are transactions left to follow, it picks the first of these, adds it to the set, and looks at its previous hops according to $\fprev$ (line~6).  It then adds these previous hops to the set of transactions (line~9) and continues.  Again, this algorithm contains an additional check in the case of the validation heuristic (line~7), which we describe in Section~\ref{sec:validation}.

\begin{algorithm}[t!]
\SetAlgoLined
\KwResult{ $\backwardtxs{\txstart,\heuristic}$ }
\SetKwInOut{Input}{Input}
\Input{ $\txstart$, $\heuristic$, $\clustertxs$, $\changestrategy$, $\clustertxfeatures$, $\clusteraddrfeatures$}

$\backwardtxs{\txstart,\heuristic} \leftarrow \emptyset$

    $\backwardscope\leftarrow \{\txstart\}$

    \While{$|\backwardscope| > 0$}{
    $\txcur \leftarrow \backwardscope[0]$

    $\backwardtxs{\txstart,\heuristic} \gets\backwardtxs{\txstart,\heuristic}\cup\{\txcur\}$

    $\fprevResult \leftarrow \fprev(\txcur, \changestrategy, \clustertxfeatures, \clusteraddrfeatures)$

    \uIf{$\heuristic= \validation$}{
        $\fprevResult\gets \fprevResult \cap \clustertxs$
    }

    $\backwardscope\leftarrow \backwardscope \cup \{\fprevResult\}$
    }

\Return $\backwardtxs{\txstart,\heuristic}$
\caption{$\followBwd$}
 \label{algo:followB}
\end{algorithm}

\section{Cluster Validation}\label{sec:overall-validation}

Currently, the clusters output by the co-spend heuristic are largely treated as ground truth, despite the fact that there exist techniques such as Coinjoin that invalidate them.  In this section, we thus investigate ways to improve one's confidence in the results of this heuristic.  In particular, we explore two approaches, each of which is applicable in a different scenario.

Our first approach, described in Section~\ref{sec:dataScience}, is a classifier for co-spend clusters that attempts to distinguish between TP and FP clusters.
This type of classifier can implicitly be realized by a Coinjoin detection mechanism, such as the one implemented in BlockSci, and indeed when we implement this approach we find it achieves 87.5\% accuracy.  Our classifier, which is based on Random Forest, achieves 89.2\% accuracy.
It achieves, however, a much lower false negative rate (10\% as compared to 20\%), which in turn lowers the risk of an investigator or researcher making an incorrect assumption about the results of the co-spend heuristic.
Furthermore, our classifier is more robust as it depends on the behavior of entities in general rather than just the characteristics of a single Coinjoin transaction (which can be changed by a Coinjoin service such as JoinMarket to avoid detection).

Our second approach, described in Section~\ref{sec:validation}, links together transactions within the same co-spend cluster that our heuristics from Section~\ref{sec:chain} identify as belonging to the same peel chain.  In doing so, we increase our confidence that these transactions were indeed performed by the same entity.  Moreover, if we run it for every transaction in the cluster then we see that TP and FP clusters have different behaviors in terms of how many distinct peel chains they contain.

\subsection{Cluster classification}\label{sec:dataScience}

Based on the transaction characteristics defined in Section~\ref{sec:txFeatures},
we computed aggregated cluster-level features. For the SegWit and locktime features,
we calculated the fraction of transactions in the cluster that had this value set to true (\texttt{prop\_segwit\_enabled}
and \texttt{prop\_locktime\_enabled} respectively). For the version feature,
we calculated the proportion of version~1 transactions (\texttt{prop\_v1}).
We did not compute a feature column for version~2 transactions to avoid
multicollinearity issues, since \texttt{prop\_v2} is given by $1 - \texttt{prop\_v1}$.
Within each cluster we also determined the proportion of all available input address
types (as defined in Section~\ref{sec:addrFeats}). These values were aggregated
to a single feature using the maximal value (\texttt{address\_type\_max\_prop}).
Finally, the cluster feature (defined in
Section~\ref{sec:clusFeats}) was transformed into two classes
(\texttt{change\_strategy}): no change strategy ($\changestrategy=\text{None}$),
or an identified change strategy on either the first or last output
($\changestrategy \in\{-1, 0, 1\}$).

Due to the small sample size (60~FP and 60~TP samples), we selected
classification models that do not require extensive hyper-parameter tuning and
tend to perform very well in a default setting~\cite{Liaw2002}. We applied
Random Forest (RF)~\cite{Breiman2001,Liaw2002}, which is a popular
and powerful machine learning method.
RF is an advancement of single classification and regression trees (CART~\cite{Breiman1984}).
As compared to CART, RF can handle a large number of covariates effectively without overfitting and are able to account for correlation as well as interactions among features.
Another important property of RF is that it immediately
provides internal variable importance measures that can be used to rank covariates.
For fitting of the CART-based RF approach, we used the implementation in the
\textsf{R}-package \emph{ranger}~\cite{Wright2017}. As an alternative, we also
applied the \emph{cforest} implementation from the package
\emph{party}~\cite{Strobl2007,Strobl2008}.

\subsubsection{Classification results}

We fit RF models with 500 trees to the dataset consisting of the
features described above and using the cluster type (TP/FP) as the target variable.
First, we fit the models to the full dataset and analyzed the intrinsic
variable importance measures.  According to both the CART-based RF and the
\emph{cforest} model, the most important features were the proportion of
SegWit transactions, the proportion of version~1 transactions, and the
proportion of transactions with enabled locktime.

\begin{figure}[t]
\centering
\includegraphics[width=0.8\columnwidth]{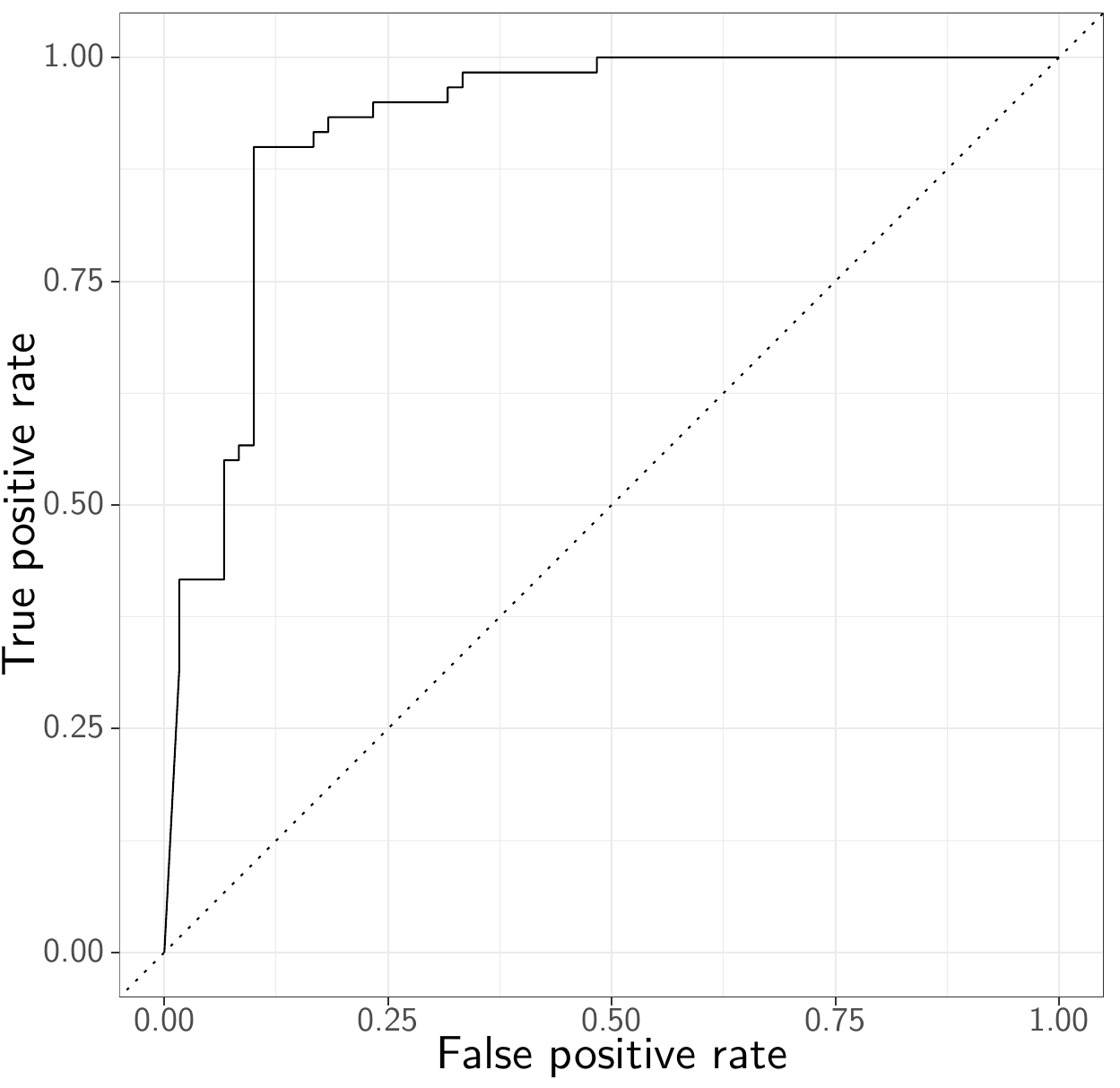}
\caption{The ROC curve for our Random Forest model (AUC=0.923).}
\label{fig:roc}
\end{figure}

For training and testing of the models we implemented a cross-validation (CV)
procedure. Accuracy, meaning the proportion of correctly classified instances,
was chosen as a model performance evaluation metric. The mean accuracy values
and their associated standard errors after a 5-fold CV are shown in
Table~\ref{tab:model-summary}, and the ROC curve is in Figure~\ref{fig:roc}.
We obtain a mean accuracy between 84\% and 89\%.
According to the standard errors, these values are also relatively stable
on the CV-testing folds.

Overall, while our classifier would of course benefit from extended experimentation with a larger dataset, these results and the high level of accuracy suggest that it would be possible to deploy this method in the manner suggested earlier in this section; i.e., for an investigator to use it to gain some confidence in the results of the co-spend heuristic at an early stage in an investigation.

\begin{table}[t]
  \centering%
\begin{tabular}{lrr}
  \toprule
Statistic & RF & Conditional RF \\
  \midrule
Mean accuracy & 0.892 & 0.842 \\
  Standard error & 0.017 & 0.031 \\
   \bottomrule
\end{tabular}
  \caption{Performance of our Random Forest model after 5-fold cross-validation.}%
  \label{tab:model-summary}%
\end{table}

\subsubsection{Comparison with BlockSci}\label{sec:coinjoin}

\begin{table}[t]
 \centering%
\begin{subtable}{0.95\columnwidth}
\centering
\begin{tabular}{rrr@{\extracolsep{5pt}}r}
\toprule
           & \multicolumn{2}{c}{Truth} & Class error  \\
\cmidrule{2-3}\cmidrule{4-4}
Prediction & FP & TP                   &              \\
FP         & 54 &  7                   & 11.5 \%      \\
TP         &  6 & 53                   & 10.2 \%      \\
\bottomrule
\end{tabular}
\caption{Confusion matrix of our RF model (summed values after 5-fold CV).}
\label{tab:rf-conf}
\end{subtable}
\vspace*{12pt}
\begin{subtable}{0.95\columnwidth}
\centering
\begin{tabular}{rrr@{\extracolsep{5pt}}r}
\toprule
           & \multicolumn{2}{c}{Truth} & Class error  \\
\cmidrule{2-3}\cmidrule{4-4}
Heuristic  & FP & TP                   &              \\
FP         &  48 &  3                   & 5.9 \%        \\
TP         &  12 &  57                   & 17.4 \%        \\
\bottomrule
\end{tabular}
\caption{Confusion matrix for the BlockSci classifier.}
\label{tab:blocksci-conf}
\end{subtable}
 \caption{Comparison of classification results.}%
 \label{tab:conf-mat}%
\end{table}

To compare our approach to the current state of the art, we explore the Coinjoin detection feature available in BlockSci~\cite{blocksci} as the function \verb#isCoinjoin#. This function works at the level of transactions, meaning given a transaction it outputs $0$ or $1$ according to whether or not it seems to be a Coinjoin, as identified by a heuristic developed by Goldfeder et al.~\cite{goldfeder2017cookie}.

Using this function, we define a cluster-level classifier by saying that if any transaction in the cluster is flagged as a Coinjoin, the entire cluster is a false positive. We tested this classifier on our full dataset of TP and FP clusters; the results are in Table~\ref{tab:blocksci-conf}. As we can see, both BlockSci and our classifier have high accuracy (87.5\% and 89.2\% respectively), with BlockSci having twice as many false negatives and our classifier having more false positives (7 as compared to 3). As argued earlier, the false negative rate is more crucial in our imagined use case, as a false negative could cause an investigator to believe that a cluster represents a single entity when in fact it does not. Furthermore, the fact that our classifier is based on all of the features within a cluster makes it more robust than our constructed BlockSci classifier, which depends only on the features of a single transaction and can thus be easily evaded by constructing Coinjoins without these specific features.

\subsection{A validation heuristic}\label{sec:validation}

Our second method for increasing the confidence we have in a cluster focuses less on the cluster as a whole and more on the connections between individual transactions.  In particular, we utilize the heuristics defined in Section~\ref{sec:chain} to partition the transactions of a cluster into peel chains.

\subsubsection{Defining the heuristic}

Our starting point is a co-spend cluster $\cluster$, represented by the tuple $(\clusteraddrs, \clustertxs)$. Next, we run $\followFwd$ and $\followFwd$ with the parameter $\heuristic = \validation$ for every $\txstart\in\clustertxs$.  Crucially, this parameter means that we do not follow any transactions that are not already in the cluster.  For a given $\txstart$ this gives us the sets $\forwardtxs{\txstart,\validation}$ and $\backwardtxs{\txstart,\validation}$, which collectively represent all transactions within the cluster that lie along the same peel chain as the starting one.  We denote the union of these two sets as $\valCluster$.

After obtaining the set $\{\valCluster\}_\txstart \in \clustertxs$ of all such peel chains, we noticed that some peel chains contained overlapping but not identical sets of transactions, according to the starting transaction $\txstart$.  We thus merged these overlapping peel chains in a transitive fashion to end up with a set of distinct peel chains, $\valClusterr$, that is a partition of all transactions in the cluster.  To measure the overall tendency for a cluster to form peel chains, we use the value $\vall = \frac{\arrowvert \valClusterr \arrowvert}{\arrowvert \clustertxs \arrowvert}$, which is closer to $1$ if a cluster consists of many peel chains and closer to $0$ if it consists of fewer.

Our \emph{validation heuristic} then says that for any two transactions $\tx_1,\tx_2\in\mathsf{Pchain}_\mathsf{V}$ for $\mathsf{Pchain}_\mathsf{V}\in\valClusterr$ (i.e., two transactions that are part of the same cluster and part of the same peel chain), we can have higher confidence that they were performed by the same entity than for two transactions $\tx_1,\tx_2\in\clustertxs$.

\subsubsection{Applying the heuristic}

It is not possible to assess the accuracy of our validation heuristic directly, as we do not have the relevant ground-truth data.  For our FP clusters, for example, we knew that they contained at least one Coinjoin but did not have any information about the other transactions (and indeed for some FP clusters it was clear they contained other Coinjoins beyond the ones we were given).  For our TP clusters, we knew that all transactions were performed by the same entity but not if they represented the same peel chain.

\begin{figure}[t]
\centering
\captionsetup[subfigure]{skip=-5pt}
\begin{subfigure}{\columnwidth}
\includegraphics[width=\columnwidth]{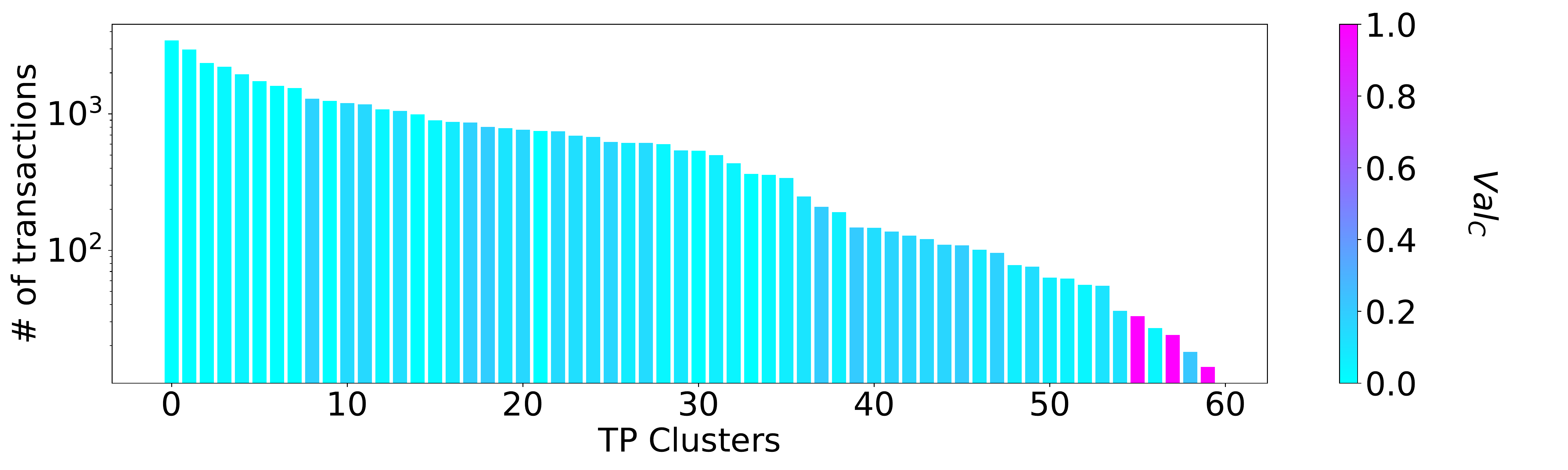}
  \label{fig:valid}
  \caption{TP clusters}
\end{subfigure}
\begin{subfigure}{\columnwidth}
\includegraphics[width=\columnwidth]{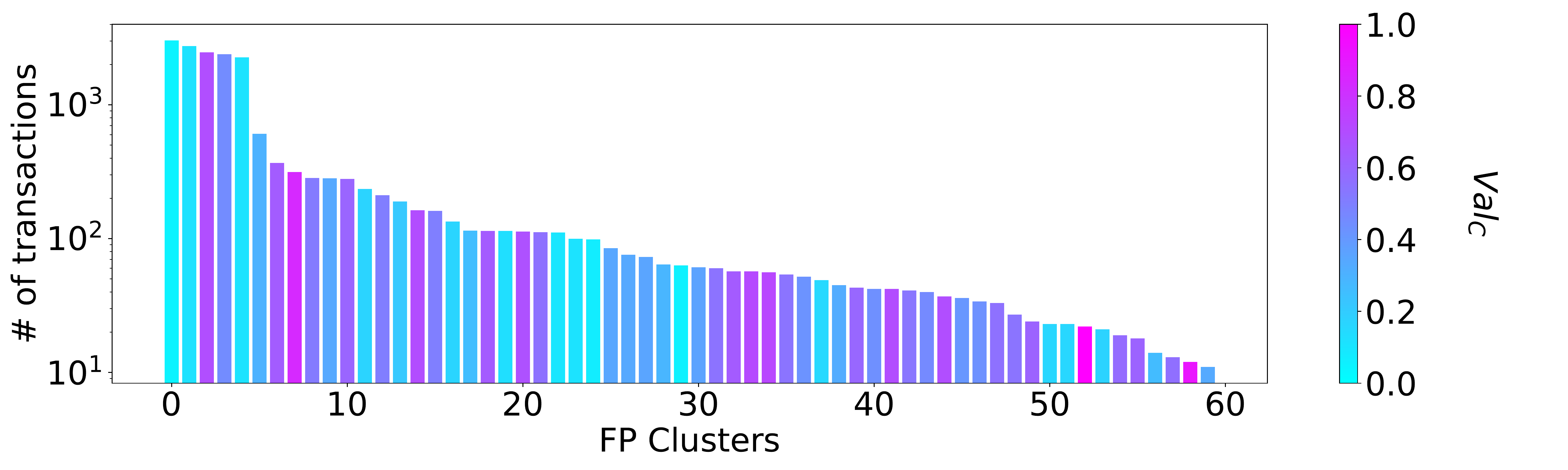}
  \label{fig:valid2}
  \caption{FP clusters}
\end{subfigure}
\caption{For our 60 TP and FP clusters, the distribution of $\arrowvert \vall \arrowvert$ (the color of the bar), with the clusters ordered from left to right by $\arrowvert \clustertxs\arrowvert$ (the height of the bar, on a log scale.}
\label{fig:overall-valid}
\end{figure}

Instead, we used our validation heuristic to understand the behavior of our TP and FP clusters.
To this end, we ran the validation heuristic for each of our clusters and looked at the resulting value of $\vall$.  As our results in Figure~\ref{fig:overall-valid} show, the FP clusters had significantly higher values of $\vall$ on average: 0.43 as compared to 0.14 for TP clusters.
Overall, $\vall$ did not increase with the size of the cluster, despite the possible expectation that clusters with a higher number of transactions would form a higher number of peel chains.  This suggests instead that bigger clusters tend to be more predictable in terms of their behavior, which is perhaps not surprising if we consider that the operators of these big clusters use automated scripts in order to form their transactions.

In terms of using this heuristic in practice, there are currently several \emph{nested services}~\cite{nested-services} that operate using accounts maintained at a variety of different exchanges.  Using just the co-spend heuristic, these nested services would thus appear to be operated by the same entity as the exchange they use.  Using our validation heuristic, however, it would be possible to separate out the activities of these nested services (which would likely form their own peel chains) from the activities of the exchange itself.

Furthermore, while we defined our validation heuristic for pairs of transactions, as we can see in Figure~\ref{fig:overall-valid} a lower value of $\vall$ can also increase our confidence in the cluster overall.  This is difficult to quantify given our current source of ground truth, however, so we leave as open work a more thorough evaluation of the impact of this heuristic on overall cluster confidence.

\section{Expanding Clusters}\label{sec:tracingHeuristic}

Our expansion heuristic is structurally similar to our validation heuristic, in that it is also based on the ability to identify peel chains, but it has a different goal: to identify new transactions that were not already in the cluster but for which we nevertheless have high confidence that they were formed by the same entity.  In this approach, this heuristic more closely resembles previous change heuristics in the literature.

\subsection{Defining the heuristic}

As with the validation heuristic, our starting point is a co-spend cluster $\cluster$ represented by a tuple $(\clusteraddrs, \clustertxs)$.  Next, we run $\followFwd$ and $\followBwd$ with the parameter $\heuristic = \expansion$ for every $\txstart\in\clustertxs$.  This gives us the sets $\forwardtxs{\txstart,\expansion}$ and $\backwardtxs{\txstart,\expansion}$ for every $\txstart$, which still represent the set of all transactions that lie along the same peel chain as $\txstart$ but crucially may contain transactions that are not already in the cluster.  We denote the union of these sets, representing all identified transactions, as $\mathsf{Txs}_\expansion$ and denote by $\expandcluster$ the set of newly identified transactions; i.e., the ones that weren't already in the cluster ($\mathsf{Txs}_\expansion \setminus \clustertxs$).
Our \emph{expansion} heuristic then says that all transactions in $\expandcluster$ were carried out by the same entity represented by the cluster.

After defining this heuristic, our goal was to identify its accuracy and effectiveness.  To measure effectiveness we defined the \emph{expansion factor} $\expansfactor$ as the increase of a cluster's coverage in terms of its number of transactions ($100 \cdot \frac{|\expandcluster|}{|\clustertxs|}$).  %
To measure accuracy, we treated as ground truth the set of tags provided to us in the Chainalysis Reactor tool,\footnote{\url{https://www.chainalysis.com/chainalysis-reactor/}}
which are tags that are gathered internally by Chainalysis and from public websites and documents.
In particular, for each transaction in $\expandcluster$, if Chainalysis had tagged it as belonging to an entity then we considered it a false positive.  If it had no tag for the transaction, we considered it an \emph{unknown positive}; i.e., we could not be sure that the transaction was formed by the same entity, but there was at least no evidence to the contrary.  We then considered the \emph{false discovery rate} $\accuracy$ as the number of false positives divided by the size of $\expandcluster$ (which is the standard definition for false discovery rate if we treat unknown positives as true positives).

In running the heuristic in its basic form, however, we encountered two problems.  First, because previous transactions were not filtered out in $\followBwd$ in the way they were for the validation heuristic, the set $\backwardscope$ was significantly larger and it became computationally infeasible to run the algorithm for longer peel chains.  Second, including so many transactions also increased the possibility of encountering a false positive, as discussed in Section~\ref{sec:one-hop}, and thus made the algorithm more prone to error.  For both of these reasons we decided to limit our expansion heuristic and only follow peel chains forwards using $\followFwd$.

\begin{figure}[t]
\centering
\begin{subfigure}{0.4\textwidth}
  \includegraphics[width=\linewidth]{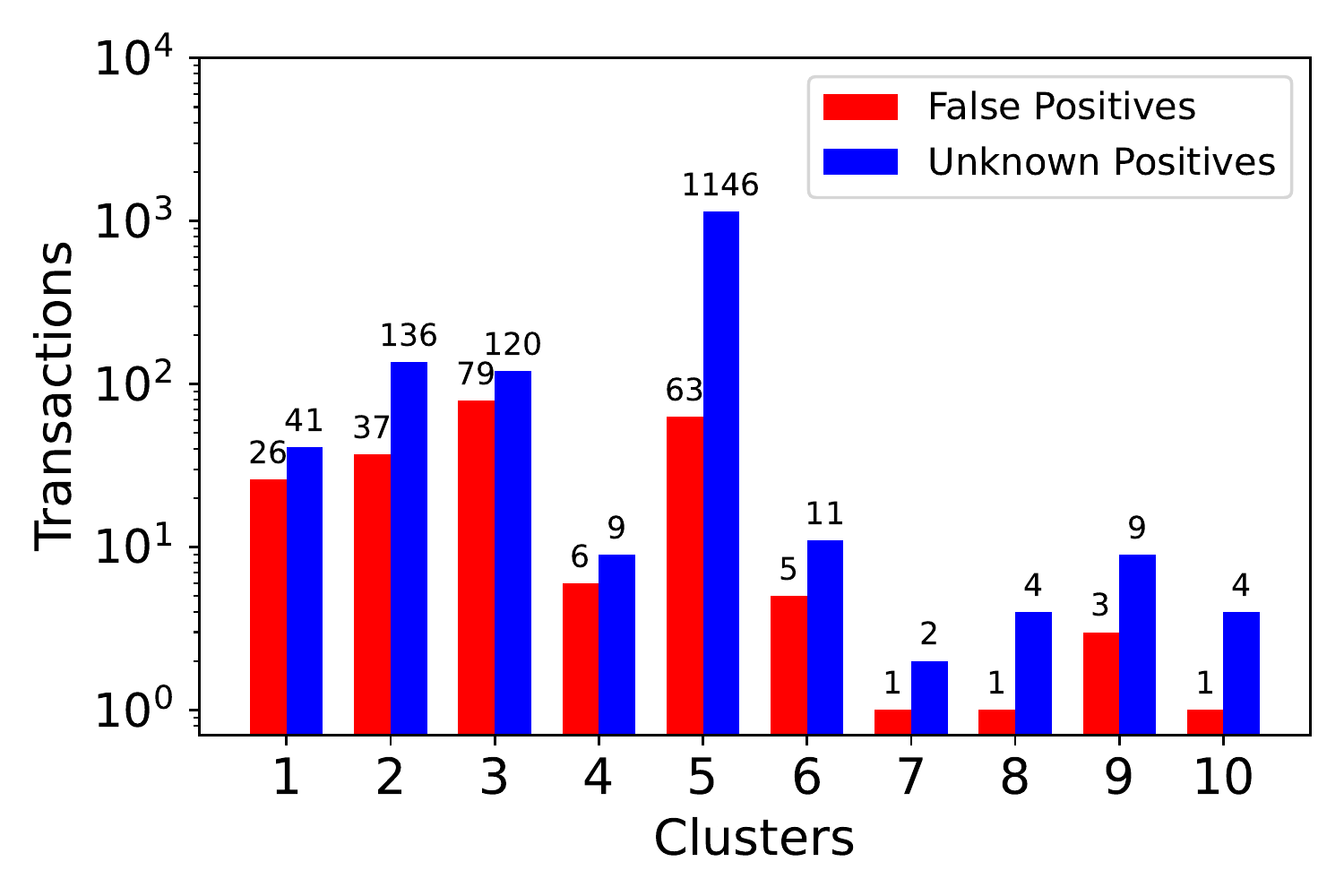}
  \caption{$\fnext$}
  \label{fig:expansion1}
\end{subfigure}
\begin{subfigure}{0.4\textwidth}
  \includegraphics[width=\linewidth]{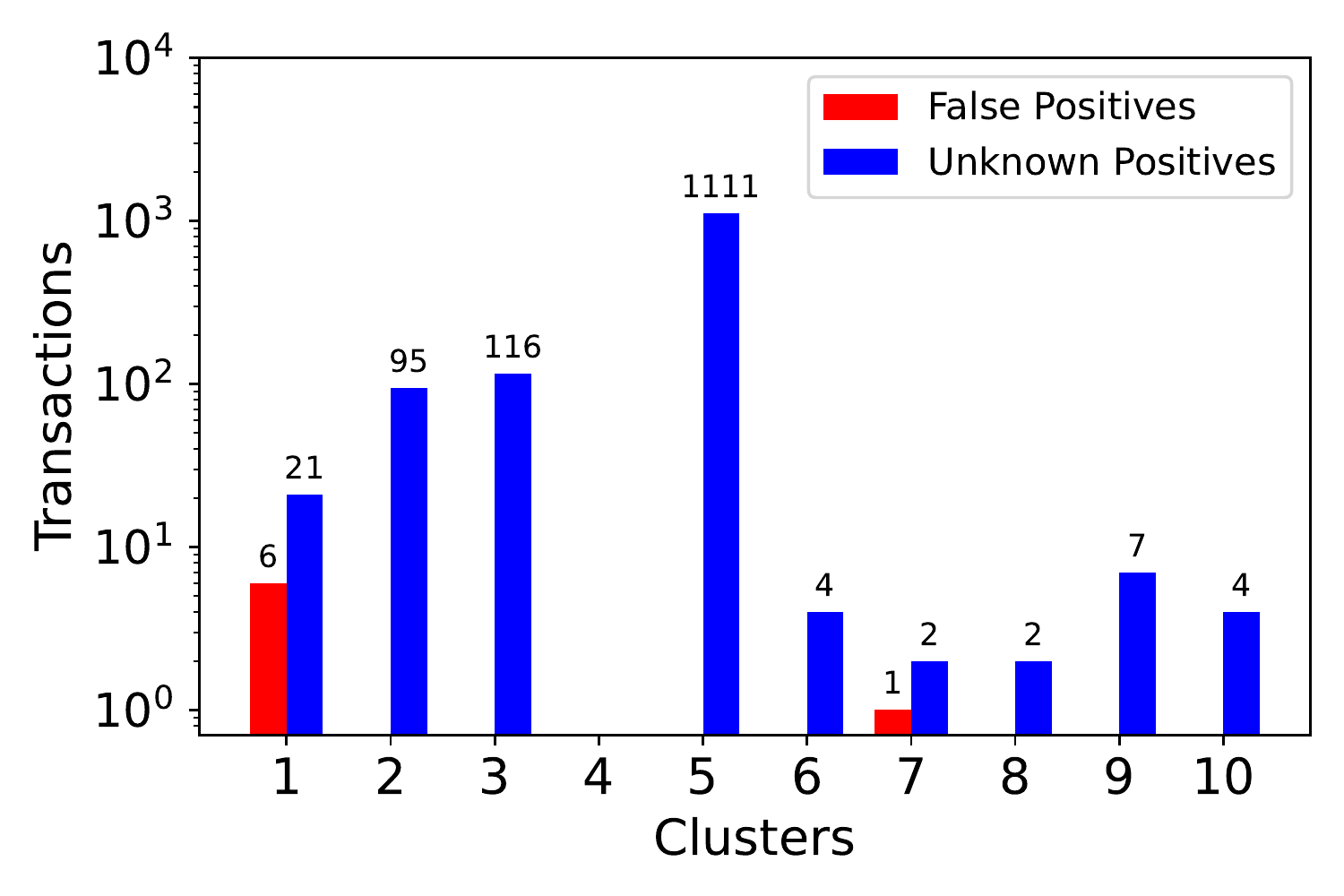}
  \caption{$\modfnext$}
  \label{fig:expansion2}
\end{subfigure}
\caption{Evaluation of $\fnext$ and the modified algorithm $\modfnext$ for the ten TP clusters with the initial highest $\accuracy$, with the number of transactions on a log scale.}
\end{figure}

After running this version of our heuristic, we achieved an $\accuracy$ of $0.62\%$, which was already quite low relative to the other heuristics (as we see below in Section~\ref{sec:expansion-results}).  Nevertheless, after manually inspecting some of the false positive transactions, we observed that the main cause was following outputs to transactions with multiple inputs that were in turn part of a bigger cluster.
We thus added an extra condition into $\fnext$ requiring that the inputs in the next hop $\txnext$ represented the entirety of the addresses in their cluster.
In other words, we added only those transactions whose inputs were only ever co-spent with each other.
We call this modified change heuristic $\modfnext$.

To illustrate the effect of this modification, Figure~\ref{fig:expansion1} shows the ten TP clusters for which $\fnext$ had the highest $\accuracy$ and thus performed the least well.  We then re-ran expansion using $\fnext2$; Figure~\ref{fig:expansion2} shows the results for the same ten clusters.  As we can see, in eight of the clusters, $\modfnext$ eliminated all false positives, at the cost of missing a relatively small number of unknown positives.

\subsection{Evaluating the heuristic}\label{sec:expansion-results}

In order to best evaluate our expansion heuristic, we sought to compare it with previous change heuristics.

\begin{sarahlist}

\item[Androulaki et al.~\cite{Androulaki2013b}] identify the change output in a transaction $\tx$ if (1) the transaction has exactly two outputs, and (2) it has the only \emph{fresh} address in $\tx.\outp$, meaning $\txoutputAddr$ is the only one appearing for the first time in the blockchain.

\item[Meiklejohn et al.~\cite{Meiklejohn2013}] identify the change output in a transaction $\tx$ if (1) it has the only fresh address in $\tx.\outp$; (2) $\tx$ is not a coin generation; and (3) there is no \emph{self-change address} in $\tx.\outp$, meaning no address used as both an input and an output.

\item[Goldfeder et al.~\cite{goldfeder2017cookie}] use the same conditions as the one by Meiklejohn et al.\ but additionally require that (4) the transaction $\tx$ is not a Coinjoin.

\item[Ermilov et al.~\cite{Ermilov2018}] were the first to consider not only the behavior of the outputs and their addresses but also the value they received.  They identify the change output in a transaction $\tx$ if (1) the transaction has exactly two outputs; (2) the transaction does not have two inputs; (3) there is no self-change address; (4) the output has the only fresh address in $\tx.\outp$; and (5) the output's value is significant to at least the fourth decimal place.

\end{sarahlist}

We implemented each of these heuristics and ran the expansion heuristic on each of our 60 TP clusters using these algorithms as well as our own algorithms $\fnext$ and $\modfnext$.  The results, in terms of false discovery rate ($\accuracy$) and expansion factor ($\expansfactor$), are in Table~\ref{tab:ExpansionResults}.

\begin{table}[t]
\centering
\begin{tabular}{l S[table-format=3.2] S[table-format=2.2]}
  \toprule
Heuristic & {$\expansfactor$} & {$\accuracy$} \\
  \midrule
$\fnext$   &  147.43   & 0.62  \\
$\fnext 2$ &  124.46   & 0.02  \\
Androulaki et al.~\cite{Androulaki2013b} &  93.03    & 64.19  \\
Meiklejohn et al.~\cite{Meiklejohn2013} &  79.94    & 51.64  \\
Goldfeder et al.~\cite{goldfeder2017cookie} &  73.7   & 48.7  \\
Ermilov et al.~\cite{Ermilov2018} &  28.6    & 12.7  \\
\bottomrule
\end{tabular}
\caption{The expansion factor and false discovery rate of $\fnext$ and $\modfnext$, as evaluated on our 60 TP clusters and as compared with previous change heuristics.  Both metrics are averaged across all clusters.}
\label{tab:ExpansionResults}
\end{table}

As Table~\ref{tab:ExpansionResults} shows, our heuristics achieve both a significantly lower false positive rate than all previous heuristics and a significantly higher expansion rate. The heuristics from Androulaki et al.\ and Meiklejohn et al.\ have the highest false discovery rates, which is somewhat expected given that Bitcoin has changed considerably since they were introduced in 2013.  As might also be expected, the heuristic from Goldfeder et al.\ achieved similar results to the one from Meiklejohn et al., with the extra Coinjoin requirement reducing both the expansion and the false positive rates by a small amount.  Finally, Ermilov et al.\ achieved the lowest $\accuracy$ of the four because of the stricter conditions of their heuristic, but this came at the expense of having the lowest expansion rate.

\subsection{Case studies}
\label{sec:case-studies}

To test our expansion heuristic in practice, we sought to run it for clusters
that had been associated with known illicit activities.

\subsubsection{Bitcoin Fog}\label{sec:btcfog}

\begin{figure*}[t]
\centering
\includegraphics[width=0.8\textwidth]{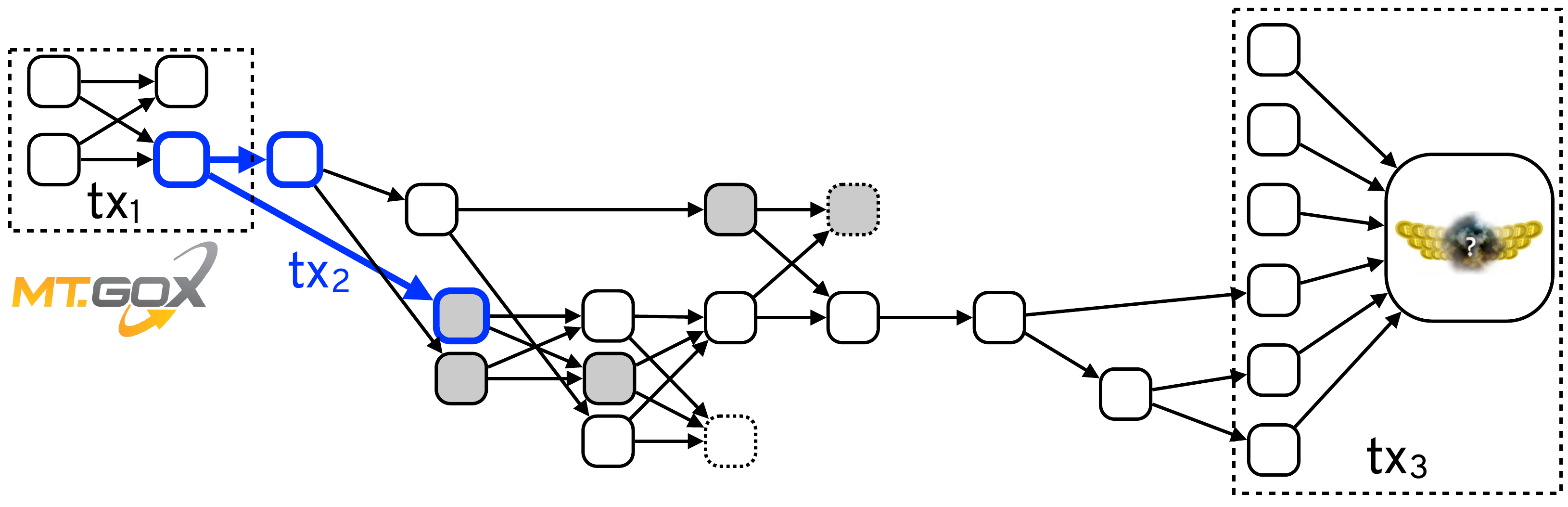}
\caption{Transactions representing the link between a withdrawal from Mt.\ Gox ($\tx_1$) and a deposit into Bitcoin Fog ($\tx_3$), with an additional transaction of interest, $\tx_2$, highlighted in blue.  The gray nodes represent multiple UTXOs of the same address and transaction outputs with a dashed outline are unspent.}
\label{fig:btcfog}
\end{figure*}

To start, we looked at a recent case against Roman Sterlingov, who was accused of being the operator of the Bitcoin Fog mixing service for almost 10 years~\cite{bitcoinfog-wired}.  According to the affidavit of an IRS special agent~\cite{bitcoinfog-affidavit}, one of the main pieces of evidence against Sterlingov was the connection of a deposit made in 2011 to the Bitcoin Fog cluster ($\tx_3$ in \Cref{fig:btcfog}) with a withdrawal from the Mt.\ Gox exchange cluster ($\tx_1$), in which Sterlingov had an account using his real name.

We first checked whether or not the cluster that received the coins from Mt.\ Gox was the same as the cluster that sent the coins to Bitcoin Fog, in order to check if it would be possible to link $\tx_1$ with $\tx_3$ based solely on the co-spend heuristic.  This was not the case, however, meaning it was necessary to take intermediate transactions into account.

We then followed the funds from the Mt.\ Gox withdrawal forwards, using $\followFwd$, to see if we would reach the deposit to Bitcoin Fog.  Both $\fnext$ and $\modfnext$ failed after only one hop, however, as the two outputs in $\tx_2$ had the same address features and were spent in transactions with the same features.  Our algorithms were thus unable to isolate the change output.  These outputs were both furthermore \emph{fresh}, meaning it was their first appearance in the blockchain, so the other change heuristics described in Section~\ref{sec:expansion-results} also would have been unable to follow the transaction forwards.

We thus worked backwards instead; i.e., we started with the deposit to Bitcoin Fog and followed the funds backwards to see if we would eventually find the withdrawal from Mt.\ Gox.  While running $\followBwd$ for all transactions in a cluster was computationally infeasible, it was possible to do it here as we started with only a single transaction.  Indeed, after seven hops, we ended up with $\tx_1$ in our set $\backwardtxs{\tx_3,\expansion}$.  While this same analysis was likely done manually by the IRS agent, this would quickly become infeasible if there were more intermediate hops; furthermore, manual analysis is arguably more error-prone as it is subject to human judgment.

While successful in this case, it is important to remember our discussion in Section~\ref{sec:one-hop} that $\followBwd$ is more prone to false positives than $\followFwd$.  Indeed, in 2011 all transactions had the same features (as the ones we use were not introduced until years later), meaning $\followBwd$ could continue backwards indefinitely without ever registering a change in the entity performing the transactions.  In our case, the paths from the last three deposits in $\tx_3$ all led back to the same origin (the second output in $\tx_1$), which in turn sent all its money to these three inputs and two unspent UTXOs, meaning we could be more sure in the link between the two.
While care must thus be taken when using $\followBwd$ in this way, applying it to a present-day scenario would likely be more safe as transactions would be expected to have a more diverse set of features.

\subsubsection{Tracking ransomware addresses}
\label{sec:expansion-usecases}

We next looked at a broader set of addresses associated with ransomware.
On December 22, 2021 we scraped the `Ransomware Help \& Tech
Support` forum of Bleeping
Computer,\footnote{\url{https://www.bleepingcomputer.com/forums/f/239/ransomware-help-tech-support/}}
a well known and actively used resource. We extracted Bitcoin addresses from
627 posts between October 2013 and December 2021.  Of the addresses, 410 were distinct and
213 of these were never used, indicating that those victims did not pay the
ransom.

For the remaining 197 addresses, we began by identifying their co-spend
clusters.  For 75 addresses, this cluster was a singleton, meaning it was a
one-time address.  For 102 addresses, the cluster was relatively small (58
addresses on average).
For 20 addresses, the resulting cluster was very large, with an
average size of 2.9 million addresses.  This suggested that these addresses
belonged to a large service rather than an individual, indicating that those
ransomware operators used a custodial solution (such as an exchange) rather than
run their own wallet.  To confirm this, we obtained the category of the tags
for these clusters from Chainalysis and found that they all belonged to either
an exchange, a hosted wallet service, a mixing service, or a darknet market.
We thus excluded these clusters from our analysis.

The remaining 177 addresses were associated with 52 different
ransomware families, with Dharma and Xorist appearing the most frequently
(with 38 and 11 addresses respectively), and formed 169 distinct co-spend
clusters.  For each
of these clusters we applied and evaluated our expansion heuristic (using
$\modfnext$).  Across all clusters, the average expansion factor was 257.5,
with a maximum expansion factor of 3400 for a CrypMIC cluster.
There were 32 clusters that did not expand at all, belonging to 20
different ransomware families; 23 of these were singleton clusters.
In addition to expanding the clusters, for each hop we followed we also
collected the \emph{counterparty} addresses; i.e., the non-change output
addresses that represented the entity or entities to whom the peel chain
operator sent coins.  In total we collected 91,493 counterparty addresses,
88,518 of which were distinct.  The vast majority (96\%) of these counterparty addresses were one-time addresses.  This is consistent with them belonging to services like exchanges, which typically provide one-time deposit addresses, but does not clearly imply this as there are many other reasons why one-time addresses occur.

While the original co-spend clusters were all distinct in terms of their associated ransomware family, after performing the expansion heuristic the clusters of two ransomware families (and only two) merged: Dharma and Phobos.  When we analyzed the counterparty addresses of these clusters as well, we observed that one of the addresses in the expanded Phobos cluster was also one of the counterparty addresses for a Dharma cluster;
i.e., an address that our expansion heuristic tagged as belonging to Phobos was also tagged as a counterparty in a peel chain formed by the Dharma ransomware operator.  In fact, this address was a counterparty 83 times, making it the most frequently used counterparty across all Dharma peel chains.  Furthermore, in several of the hops in this peel chain both transaction outputs were fresh, meaning they could not have been followed by any previous change heuristic.
While further investigation is of course needed, this is in line with the hypothesis that these ransomware families are operated by the same
entity~\cite{dharma-phobos}.

\section{Limitations and Countermeasures}
\label{sec:limitations}

Both our heuristics and our classifier could be made significantly less
effective by a motivated party randomizing the features of their transactions
and addresses.  Performing this randomization requires
a relatively high level of technological sophistication and familiarity with
Bitcoin but does not incur any computational or monetary costs, and would be
effective in making it either impossible to follow hops in a peel chain
(because their features would be randomized and thus not match the expected
behavior) or make the features associated with a cluster so varied that any
attempt to follow its transactions would be likely to result in a false
positive.
Transactions that are private and internal to the ledger of some large entity
(like an exchange) would also make it impossible for our heuristics, which rely on
public blockchain data, to work.  Given this, investigators should not rely on
the output of either our heuristics or our classifier as definitive evidence.

All previous clustering heuristics share these limitations, however, and are
still broadly effective in practice due to the fact that many entities
lack either the motivation or the technical ability to evade them.  For example,
exchanges and other regulated entities have no reason to evade these heuristics,
and previous research has shown that the ability to identify their transactions
has a knock-on effect in terms of anonymity even for participants who do actively
try to perform such evasions.

Just as with previous clustering heuristics, the heuristics presented here may
also become less effective as Bitcoin evolves.  As discussed in Section~\ref{sec:one-hop}, our heuristics produce few false positives due to the fact that many entities are consistent in their behavior and are largely non-overlapping in terms of their features.  If all entities had the same features then our heuristics would stop working, and similarly if many entities had the same features and were less consistent in their behavior then our heuristics might produce more false positives.  As a concrete example, a recent pull request in the main Bitcoin
repository\footnote{\url{https://github.com/bitcoin/bitcoin/pull/23789}}
ensures that the type of the change address matches exactly the type of the counterparty address.  If this version
of the wallet software were adopted by every entity in the Bitcoin ecosystem,
it would make our address heuristic entirely ineffective (although our
transaction and change heuristics would still work).

\section{Conclusion}\label{sec:conclusion}

In this paper, we presented heuristics to expand and validate the applicability of widely used Bitcoin clustering heuristics, using a balanced ground-truth dataset to evaluate them.
While this research arguably further reduces anonymity
in Bitcoin, we believe that it ultimately benefits the developers and users of this project in revealing the extent to which tracking flows of bitcoins is possible and motivating further research into improved anonymity protocols.  As an immediate countermeasure, users who are concerned about their privacy can switch to more privacy-focused cryptocurrencies such as Zcash and Monero, although previous work has shown that even these are subject to some degree of de-anonymization~\cite{zcashanon,moneroanon, Amritmonero,yu-monero}.

\section*{Acknowledgements}

We are grateful to Chainalysis for working with us to create the dataset that
made this work possible and to Jacob Illum and Peter Sebastian Nordholt in
particular for many helpful discussions.  We would also like to thank the
anonymous reviewers and our shepherd, Anita Nikolich, for their feedback.  
The authors were supported in part by the EU H2020 TITANIUM project under
grant agreement number 740558, in part by the Austrian
security research programme KIRAS of the Federal Ministry of Agriculture,
Regions and Tourism (BMLRT) under the project KRYPTOMONITOR (879686), and in
part by IC3 industry partners.


\begin{thebibliography}{10}

\bibitem{abramova21chi}
S.~Abramova, A.~Voskobojnikov, K.~Beznosov, and R.~B{\"o}hme.
\newblock Bits under the mattress: Understanding different risk perceptions and
  security behaviors of crypto-asset users.
\newblock In {\em CHI Conference on Human Factors in Computing Systems (CHI)},
  2021.

\bibitem{Androulaki2013b}
E.~Androulaki, G.~O. Karame, M.~Roeschlin, T.~Scherer, and S.~Capkun.
\newblock {Evaluating user privacy in Bitcoin}.
\newblock In {\em International Conference on Financial Cryptography and Data
  Security}, volume 7859 LNCS, pages 34--51, 2013.

\bibitem{Bartoletti2018}
M.~Bartoletti, B.~Pes, and S.~Serusi.
\newblock {Data mining for detecting Bitcoin Ponzi schemes}, 2018.
\newblock \url{http://arxiv.org/abs/1803.00646}.

\bibitem{bitcoinfog-affidavit}
D.~Beckett.
\newblock Statement of facts, Apr. 2021.
\newblock
  \url{https://storage.courtlistener.com/recap/gov.uscourts.dcd.230456/gov.uscourts.dcd.230456.1.1_1.pdf}.

\bibitem{biryukov2014bitcoin}
A.~Biryukov, D.~Khovratovich, and I.~Pustogarov.
\newblock Deanonymisation of clients in {Bitcoin} {P2P} network.
\newblock In {\em Proceedings of ACM CCS}, 2014.

\bibitem{Breiman2001}
L.~Breiman.
\newblock Random forests.
\newblock {\em Machine Learning}, 45(1):5--32, 2001.

\bibitem{Breiman1984}
L.~Breiman, J.~Friedman, C.~J. Stone, and R.~A. Olshen.
\newblock {\em Classification and Regression Trees}.
\newblock Chapman and Hall/CRC, 1984.

\bibitem{nested-services}
{Chainalysis Team}.
\newblock 270 service deposit addresses drive 55\% of money laundering in
  cryptocurrency, Feb. 2021.
\newblock
  \url{https://blog.chainalysis.com/reports/cryptocurrency-money-laundering-2021}.

\bibitem{wannacry-tracking}
J.~Dunietz.
\newblock {The Imperfect Crime: How the WannaCry Hackers Could Get Nabbed},
  Aug. 2017.
\newblock
  \url{https://www.scientificamerican.com/article/the-imperfect-crime-how-the-wannacry-hackers-could-get-nabbed/}.

\bibitem{Ermilov2018}
D.~Ermilov, M.~Panov, and Y.~Yanovich.
\newblock {Automatic Bitcoin address clustering}.
\newblock In {\em Proceedings of the 16th IEEE International Conference on
  Machine Learning and Applications (ICMLA 2017)}, pages 461--466, 2018.

\bibitem{wsj-hamas}
B.~Faucon, I.~Talley, and S.~Said.
\newblock {Israel-Gaza Conflict Spurs Bitcoin Donations to Hamas}, June 2021.
\newblock
  \url{https://www.wsj.com/articles/israel-gaza-conflict-spurs-bitcoin-donations-to-hamas-11622633400}.

\bibitem{bip68}
M.~Friedenbach, BtcDrak, N.~Dorier, and kinoshitajona.
\newblock {BIP 68: Relative lock-time using consensus-enforced sequence
  numbers}, 2015.
\newblock \url{https://github.com/bitcoin/bips/blob/master/bip-0068.mediawiki}.

\bibitem{Frowis2019e}
M.~Fr{\"{o}}wis, T.~Gottschalk, B.~Haslhofer, C.~R{\"{u}}ckert, and P.~Pesch.
\newblock {Safeguarding the Evidential Value of Forensic Cryptocurrency
  Investigations}, 2019.
\newblock \url{http://arxiv.org/abs/1906.12221}.

\bibitem{goldfeder2017cookie}
S.~Goldfeder, H.~Kalodner, D.~Reisman, and A.~Narayanan.
\newblock When the cookie meets the blockchain: Privacy risks of web payments
  via cryptocurrencies.
\newblock {\em arXiv preprint arXiv:1708.04748}, 2017.

\bibitem{silkroad-tracking}
A.~Greenberg.
\newblock {Prosecutors Trace \$13.4M in Bitcoins From the Silk Road to
  Ulbricht's Laptop}, Jan. 2015.
\newblock
  \url{https://www.wired.com/2015/01/prosecutors-trace-13-4-million-bitcoins-silk-road-ulbrichts-laptop/}.

\bibitem{bitcoinfog-wired}
A.~Greenberg.
\newblock {Feds Arrest an Alleged \$336M Bitcoin-Laundering Kingpin}, Apr.
  2021.
\newblock
  \url{https://www.wired.com/story/bitcoin-fog-dark-web-cryptocurrency-arrest/}.

\bibitem{bip0125}
D.~A. Harding and P.~Todd.
\newblock {BIP 125: Opt-in Full Replace-by-Fee Signaling}, 2015.
\newblock \url{https://github.com/bitcoin/bips/blob/master/bip-0125.mediawiki}.

\bibitem{harlev2018breaking}
M.~A. Harlev, H.~Sun~Yin, K.~C. Langenheldt, R.~Mukkamala, and R.~Vatrapu.
\newblock Breaking bad: De-anonymising entity types on the bitcoin blockchain
  using supervised machine learning.
\newblock In {\em Proceedings of the 51st Hawaii International Conference on
  System Sciences}, 2018.

\bibitem{Harrigan2017b}
M.~Harrigan and C.~Fretter.
\newblock {The Unreasonable Effectiveness of Address Clustering}.
\newblock In {\em Proceedings of the 13th IEEE International Conference on
  Ubiquitous Intelligence and Computing}, pages 368--373, 2017.

\bibitem{8418627}
D.~Y. Huang, M.~M. Aliapoulios, V.~G. Li, L.~Invernizzi, E.~Bursztein,
  K.~McRoberts, J.~Levin, K.~Levchenko, A.~C. Snoeren, and D.~McCoy.
\newblock Tracking ransomware end-to-end.
\newblock In {\em 2018 IEEE Symposium on Security and Privacy (SP)}, pages
  618--631, 2018.

\bibitem{huang2014botcoin}
D.~Y. Huang, H.~Dharmdasani, S.~Meiklejohn, V.~Dave, C.~Grier, D.~McCoy,
  S.~Savage, N.~Weaver, A.~C. Snoeren, and K.~Levchenko.
\newblock Botcoin: Monetizing stolen cycles.
\newblock In {\em NDSS}. Citeseer, 2014.

\bibitem{Jourdan2018}
M.~Jourdan, S.~Blandin, L.~Wynter, and P.~Deshpande.
\newblock {Characterizing Entities in the Bitcoin Blockchain}, 2018.
\newblock \url{http://arxiv.org/abs/1810.11956}.

\bibitem{blocksci}
H.~Kalodner, M.~M{\"o}ser, K.~Lee, S.~Goldfeder, M.~Plattner, A.~Chator, and
  A.~Narayanan.
\newblock Blocksci: Design and applications of a blockchain analysis platform.
\newblock In {\em 29th {USENIX} Security Symposium ({USENIX} Security 20)},
  pages 2721--2738. {USENIX} Association, Aug. 2020.

\bibitem{zcashanon}
G.~Kappos, H.~Yousaf, M.~Maller, and S.~Meiklejohn.
\newblock An empirical analysis of anonymity in zcash.
\newblock In {\em 27th {USENIX} Security Symposium ({USENIX} Security 18)},
  pages 463--477, Baltimore, MD, Aug. 2018. {USENIX} Association.

\bibitem{koshy2014bitcoin}
P.~Koshy, D.~Koshy, and P.~McDaniel.
\newblock An analysis of anoymity in {Bitcoin} using {P2P} network traffic.
\newblock In {\em International Conference on Financial Cryptography and Data
  Security (FC)}, 2014.

\bibitem{krombholz16fc}
K.~Krombholz, A.~Judmayer, M.~Gusenbauer, and E.~Weippl.
\newblock The other side of the coin: User experiences with {Bitcoin} security
  and privacy.
\newblock In {\em International Conference on Financial Cryptography and Data
  Security}, 2016.

\bibitem{Amritmonero}
A.~Kumar, C.~Fischer, S.~Tople, and P.~Saxena.
\newblock A traceability analysis of {Monero}'s blockchain.
\newblock In S.~N. Foley, D.~Gollmann, and E.~Snekkenes, editors, {\em Computer
  Security -- ESORICS 2017}, pages 153--173, Cham, 2017. Springer International
  Publishing.

\bibitem{Liaw2002}
A.~Liaw and M.~Wiener.
\newblock Classification and regression by \texttt{randomForest}.
\newblock {\em R News}, 2(3):18--22, 2002.

\bibitem{bip141}
E.~Lombrozo, J.~Lau, and P.~Wuille.
\newblock {BIP 141: Segregated Witness (Consensus layer)}, 2015.
\newblock \url{https://github.com/bitcoin/bips/blob/master/bip-0141.mediawiki}.

\bibitem{mai20soups}
A.~Mai, K.~Pfeffer, M.~Gusenbauer, E.~Weippl, and K.~Krombholz.
\newblock User mental models of cryptocurrency systems - a grounded theory
  approach.
\newblock In {\em Proceedings of the Symposium on Usable Privacy and Security
  (SOUPS)}, 2020.

\bibitem{Meiklejohn2013}
S.~Meiklejohn, M.~Pomarole, G.~Jordan, K.~Levchenko, D.~McCoy, G.~M. Voelker,
  and S.~Savage.
\newblock {A fistful of bitcoins: Characterizing payments among men with no
  names}.
\newblock In {\em Proceedings of the Internet Measurement Conference - IMC
  '13}, number~6, pages 127--140, 2013.

\bibitem{newyorker-ransomware}
R.~Monroe.
\newblock How to negotiate with ransomware hackers, June 2021.
\newblock
  \url{https://www.newyorker.com/magazine/2021/06/07/how-to-negotiate-with-ransomware-hackers}.

\bibitem{malte21}
M.~M{\"o}ser and A.~Narayanan.
\newblock Resurrecting address clustering in {Bitcoin}, 2021.
\newblock \url{https://arxiv.org/pdf/2107.05749.pdf}.

\bibitem{moneroanon}
M.~M{\"o}ser, K.~Soska, E.~Heilman, K.~Lee, H.~Heffan, S.~Srivastava, K.~Hogan,
  J.~Hennessey, A.~Miller, A.~Narayanan, and N.~Christin.
\newblock An empirical analysis of linkability in the {Monero} blockchain.
\newblock {\em Proceedings on Privacy Enhancing Technologies}, 2017.

\bibitem{whitepaper}
S.~Nakamoto.
\newblock {Bitcoin: A Peer-to-Peer Electronic Cash System}, 2008.
\newblock \url{bitcoin.org/bitcoin.pdf}.

\bibitem{Nick2015a}
J.~D. Nick.
\newblock {Data-Driven De-Anonymization in Bitcoin}.
\newblock Master's thesis, ETH Z{\"u}rich, 2015.

\bibitem{dharma-phobos}
D.~Palmer.
\newblock New {Phobos} ransomware exploits weak security to hit targets around
  the world, Jan. 2019.
\newblock
  \url{https://www.zdnet.com/article/new-phobos-ransomware-exploits-weak-security-to-hit-targets-around-the-world/}.

\bibitem{10.1093/cybsec/tyz003}
M.~Paquet-Clouston, B.~Haslhofer, and B.~Dupont.
\newblock {Ransomware payments in the Bitcoin ecosystem}.
\newblock {\em Journal of Cybersecurity}, 5(1), 05 2019.
\newblock tyz003.

\bibitem{paquet2019spams}
M.~Paquet-Clouston, M.~Romiti, B.~Haslhofer, and T.~Charvat.
\newblock Spams meet cryptocurrencies: Sextortion in the bitcoin ecosystem.
\newblock In {\em Proceedings of the 1st ACM conference on advances in
  financial technologies}, pages 76--88, 2019.

\bibitem{Pham2014}
P.~T. Pham and S.~Lee.
\newblock {Anomaly Detection in the Bitcoin System-A Network Perspective},
  2014.
\newblock \url{https://arxiv.org/abs/1611.03942}.

\bibitem{Pham2016}
T.~Pham and S.~Lee.
\newblock {Anomaly Detection in Bitcoin Network Using Unsupervised Learning
  Methods}, 2016.
\newblock \url{http://arxiv.org/abs/1611.03941}.

\bibitem{portnoff2017backpage}
R.~S. Portnoff, D.~Y. Huang, P.~Doerfler, S.~Afroz, and D.~McCoy.
\newblock Backpage and bitcoin: Uncovering human traffickers.
\newblock In {\em Proceedings of the 23rd ACM SIGKDD International Conference
  on Knowledge Discovery and Data Mining}, pages 1595--1604, 2017.

\bibitem{Ranshous2017b}
S.~Ranshous, C.~A. Joslyn, S.~Kreyling, K.~Nowak, N.~F. Samatova, C.~L. West,
  and S.~Winters.
\newblock {Exchange pattern mining in the Bitcoin transaction directed
  hypergraph}.
\newblock In {\em International Conference on Financial Cryptography and Data
  Security}, volume 10323 LNCS, pages 248--263, 2017.

\bibitem{Reid2013c}
F.~Reid and M.~Harrigan.
\newblock {An analysis of anonymity in the Bitcoin system}.
\newblock {\em Security and Privacy in Social Networks}, pages 197--223, 2013.

\bibitem{FC:RonSha13}
D.~Ron and A.~Shamir.
\newblock Quantitative analysis of the full {Bitcoin} transaction graph.
\newblock In {\em International Conference on Financial Cryptography and Data
  Security}, 2013.

\bibitem{Spagnuolo2014a}
M.~Spagnuolo, F.~Maggi, and S.~Zanero.
\newblock {BitIodine: Extracting intelligence from the Bitcoin network}.
\newblock In {\em International Conference on Financial Cryptography and Data
  Security}, volume 8437, pages 457--468, 2014.

\bibitem{Strobl2008}
C.~Strobl, A.-L. Boulesteix, T.~Kneib, T.~Augustin, and A.~Zeileis.
\newblock Conditional variable importance for random forests.
\newblock {\em BMC Bioinformatics}, 9(307), 2008.

\bibitem{Strobl2007}
C.~Strobl, A.-L. Boulesteix, A.~Zeileis, and T.~Hothorn.
\newblock Bias in random forest variable importance measures: Illustrations,
  sources and a solution.
\newblock {\em BMC Bioinformatics}, 8(25), 2007.

\bibitem{8258365}
H.~Sun~Yin and R.~Vatrapu.
\newblock A first estimation of the proportion of cybercriminal entities in the
  {Bitcoin} ecosystem using supervised machine learning.
\newblock In {\em 2017 IEEE International Conference on Big Data (Big Data)},
  pages 3690--3699, 2017.

\bibitem{child-tracking}
{United States Department of Justice}.
\newblock South {Korean} national and hundreds of others charged worldwide in
  the takedown of the largest darknet child pornography website, which was
  funded by {Bitcoin}, Oct. 2019.
\newblock
  \url{https://www.justice.gov/opa/pr/south-korean-national-and-hundreds-others-charged-worldwide-takedown-largest-darknet-child}.

\bibitem{terrorist-tracking}
{United States Department of Justice}.
\newblock Global disruption of three terror finance cyber-enabled campaigns,
  Aug. 2020.
\newblock
  \url{https://www.justice.gov/opa/pr/global-disruption-three-terror-finance-cyber-enabled-campaigns}.

\bibitem{Wright2017}
M.~N. Wright and A.~Ziegler.
\newblock \texttt{ranger}: A fast implementation of random forests for high
  dimensional data in {\Cpp} and \textsf{R}.
\newblock {\em Journal of Statistical Software}, 77(1):1--17, 2017.

\bibitem{YinRegulating}
H.~H.~S. Yin, K.~Langenheldt, M.~Harlev, R.~R. Mukkamala, and R.~Vatrapu.
\newblock Regulating cryptocurrencies: A supervised machine learning approach
  to de-anonymizing the {Bitcoin} blockchain.
\newblock {\em Journal of Management Information Systems}, 36(1):37--73, 2019.

\bibitem{236358}
H.~Yousaf, G.~Kappos, and S.~Meiklejohn.
\newblock Tracing transactions across cryptocurrency ledgers.
\newblock In {\em 28th {USENIX} Security Symposium ({USENIX} Security 19)},
  pages 837--850, Santa Clara, CA, Aug. 2019. {USENIX} Association.

\bibitem{yu-monero}
Z.~Yu, M.~H. Au, J.~Yu, R.~Yang, Q.~Xu, and W.~F. Lau.
\newblock New empirical traceability analysis of {CryptoNote}-style
  blockchains.
\newblock In I.~Goldberg and T.~Moore, editors, {\em Financial Cryptography and
  Data Security}, pages 133--149, Cham, 2019. Springer International
  Publishing.

\bibitem{Zambre2013}
D.~Zambre and A.~Shah.
\newblock {Analysis of Bitcoin Network Dataset for Fraud}.
\newblock Stanford CS 224W Project Final Report, 2013.
\newblock
  \url{http://snap.stanford.edu/class/cs224w-2013/projects2013/cs224w-030-final.pdf}.

\end{thebibliography}
\end{document}